\documentclass[journal]{IEEEtran}
\usepackage{amsmath, amsfonts, array, bm, amsthm}
\usepackage{algorithmic, algorithm}
\usepackage{subfig, textcomp, stfloats, xcolor}
\usepackage{url, verbatim, graphicx}
\usepackage{cite, breqn}
\usepackage{hhline, makecell}

\newtheorem{definition}{Definition}

\newtheorem{lemma}{Lemma}
\newtheoremstyle{iremark}
  {\topsep}   % ABOVESPACE
  {\topsep}   % BELOWSPACE
  {\upshape}  % BODYFONT
  {0pt}       % INDENT (empty value is the same as 0pt)
  {\itshape}  % HEADFONT
  {.}         % HEADPUNCT
  {5pt plus 1pt minus 1pt} % HEADSPACE
  {\thmname{#1}\thmnumber{ \itshape#2}\thmnote{ (#3)}} % CUSTOM-HEAD-SPEC

\theoremstyle{iremark}

\newtheoremstyle{noparens}%
  {}{}%
  {\itshape}{}%
  {\bfseries}{.}%
  { }%
  {\thmname{#1}\thmnumber{ #2}\mdseries\thmnote{ #3}}
\theoremstyle{noparens}
\newtheorem{lemmaNoParens}[lemma]{Lemma}

\newtheorem{theorem}{Theorem}
\newtheoremstyle{case}{}{}{}{}{}{:}{ }{}
\theoremstyle{case}

\DeclareMathOperator*{\argmin}{arg\,min}

\begin{document}
\title{Joint Design of Radar Receive Filter and Unimodular ISAC Waveform with Sidelobe Level Control}
\author{\IEEEauthorblockN{Kecheng Zhang, Ya-Feng Liu,\IEEEmembership{~Senior Member,~IEEE}, Zhongbin Wang, Weijie Yuan,\IEEEmembership{~Senior Member,~IEEE},\\Musa Furkan Keskin,\IEEEmembership{~Member,~IEEE}, Henk Wymeersch,\IEEEmembership{~Fellow,~IEEE}, Shuqiang Xia}
  % <-this % stops a space
  \thanks{This work is supported in part by the Swedish Research Council (VR) through the project 6G-PERCEF under Grant 2024-04390, in part by National Natural Science Foundation of China under Grant 62471208, in part by Guangdong Provincial Natural Science Foundation under Grant 2024A151510098, in part by ZTE Industry-University-Institute Cooperation Funds under Grant No. IA20240610003, and in part by Shenzhen Science and Technology Program under Grant JCYJ20240813094627037. The work of Y.-F. Liu was supported in part by the National Natural Science Foundation of China (NSFC) under Grant 12371314 and Grant 12021001. \emph{(Corresponding Author: Ya-Feng Liu.)}}
  \thanks{Kecheng Zhang and Weijie Yuan are with the School of Automation and Intelligent Manufacturing, Southern University of Science and Technology, Shenzhen 518055, China (e-mail: zhangkc2022@mail.sustech.edu.cn; yuanwj@sustech.edu.cn)}
  \thanks{Ya-Feng Liu is with the Ministry of Education Key Laboratory of Mathematics and Information Networks, School of Mathematical Sciences, Beijing University of Posts and Telecommunications, Beijing 100876, China (email: yafengliu@bupt.edu.cn)}
  \thanks{Musa Furkan Keskin and Henk Wymeersch are with Department of Electrical Engineering, Chalmers University of Technology, 41296 Gothenburg, Sweden (e-mail: \{furkan; henkw\}@chalmers.se)}
  \thanks{Zhongbin Wang and Shuqiang Xia are with the ZTE Corporation and the State Key Laboratory of Mobile Network and Mobile Multimedia Technology, Shenzhen 518055, China (e-mail: \{wang.zhongbin; xia.shuqiang\}@zte.com.cn)}
}



\maketitle
\begin{abstract}
  Integrated sensing and communication (ISAC) has been considered a key feature of next-generation wireless networks. This paper investigates the joint design of the radar receive filter and dual-functional transmit waveform for the multiple-input multiple-output (MIMO) ISAC system. While optimizing the mean square error (MSE) of the radar receive spatial response and maximizing the achievable rate at the communication receiver, besides the constraints of full-power radar receiving filter and unimodular transmit sequence, we control the maximum range sidelobe level, which is often overlooked in existing ISAC waveform design literature, for better radar imaging performance. To solve the formulated optimization problem with convex and nonconvex constraints, we propose an inexact augmented Lagrangian method (ALM) algorithm. For each subproblem in the proposed inexact ALM algorithm, we custom-design a block successive upper-bound minimization (BSUM) scheme with closed-form solutions for all blocks of the variable to enhance the computational efficiency. Convergence analysis shows that the proposed algorithm is guaranteed to provide a stationary and feasible solution. Extensive simulations are performed to investigate the impact of different system parameters on communication and radar imaging performance. Comparison with the existing works shows the superiority of the proposed algorithm.
\end{abstract}


\section{Introduction}\label{sec:introduction}
Integrated sensing and communication (ISAC) unify sensing and communication (S$\&$C) tasks into a single system, improving efficiency and performance by sharing various resources like spectrum and hardware \cite{ISAC1, ISAC2}. With its potential to support emerging applications requiring high-quality wireless connections and accurate sensing, such as autonomous driving and smart homes, ISAC is widely regarded as a key enabler for next-generation wireless networks \cite{ISAC5, 10596930}. However, S$\&$C subsystems have distinct performance requirements \cite{10188491}: sensing favors unimodular and deterministic waveforms, while communication relies on waveforms with high degrees of freedom (DoFs) and randomness for efficient information transmission. Unimodular waveform here means the modulus of the complex baseband signal is unity at all times. As a result, the interference between them is inevitable.

Since wireless systems often need to simultaneously serve multiple users and meet their various sensing and communication needs, multiple-input multiple-output (MIMO) technology \cite{10556753, 10500425} becomes particularly important in ISAC systems. Early ISAC research focused on the MIMO radar-communication coexistence by mitigating interference through spectrum-sharing techniques like dynamic spectrum access \cite{saruthirathanaworakun2012opportunistic} and null-space projection \cite{mahal2017spectral}. Although these approaches enabled S$\&$C coexistence, the systems were designed separately and required side-information exchange, leading to additional cooperation costs \cite{Liu2018}. As an advancement from the spectrum-sharing scheme, some works \cite{10707252, 10858773} focused on designing or implementing new antenna structures for ISAC purpose, such as the fluid antenna-aided ISAC systems and its combination with reconfigurable intelligent surface. Besides the newly appeared antenna structures, dual-functional waveform design has recently drawn a lot of interest \cite{ISAC4} for its ability to sense targets and transmit information using a single device simultaneously, eliminating the need for side-information exchange in S$\&$C cooperation \cite{Liu2018} and the change of antenna structures.

Many studies have focused on the design of dual-functional waveforms for ISAC; see \cite{ISAC1} and the references therein. However, most of these works overlook the crucial aspect of controlling the sidelobes of the correlation function. In modern communication systems such as 5G NR, the transmitted signals typically occupy a certain bandwidth \cite{3gpp.36.104}. When these wideband signals are reused for sensing in ISAC systems, the resulting range sidelobes can significantly degrade sensing performance. High sidelobe levels are undesirable because they may interfere with or even obscure the reflections from distant targets or those with small radar cross-sections (RCS), potentially leading to missed detections \cite{Liu2020}. Moreover, even when the radar receiver beampattern is well-optimized, excessive range sidelobes can still cause false alarms. Therefore, effective sidelobe control is essential in ISAC waveform design to ensure reliable sensing performance.

Additionally, it is well known that power amplifiers operate most efficiently when the input signals are unimodular [12], which further motivates the need for unimodular waveform design in ISAC systems. In this paper, we address the problem of dual-functional waveform design for monostatic downlink transmission in a MIMO-ISAC system. Our objective is to jointly maximize the achievable communication rate and minimize the mean square error (MSE) of the radar beampattern while simultaneously controlling the peak sidelobe level.

\subsection{Literature Review}
Existing dual-functional waveform design research can be roughly divided into three categories: sensing-centric, communication-centric, and joint waveform design, which will be detailed below.

Sensing-centric waveform design focuses on embedding the communication information into sensing waveforms \cite{ISAC1}. For example, the communication information can be embedded into chirp waveforms \cite{chiriyath2019novel}, spatial beampattern \cite{hassanien2015dual}, and ambiguity function \cite{yang2020dual}. While these approaches generally exhibit strong radar sensing capabilities, they often face low communication rates due to the limited number of embedded information bits. In contrast, communication-centric waveform design implements radar sensing using existing communication waveforms, such as orthogonal frequency division multiplexing (OFDM) \cite{sturm2011waveform} and orthogonal time frequency space (OTFS) \cite{10279816}. However, the sensing performance of communication-centric designs is unpredictable due to the inherent randomness of communication signals and potential distortion from a high peak-to-average power ratio.

To address the limitations of separate designs and achieve trade-offs between S$\&$C, many works have focused on joint waveform design \cite{ISAC1}. This approach constructs waveforms by solving optimization problems under various S$\&$C constraints. More specifically, in \cite{xu2021rate}, the authors jointly maximized a weighted sum rate while minimizing the radar beampattern approximation MSE, constrained by per-antenna power limits, enabling rate-splitting multiple access and interference management in MIMO-ISAC systems. In \cite{liu2021cramer}, the authors minimized the Cramér-Rao bound (CRB) for direction-of-arrival (DoA) estimation by designing the beamforming matrix under individual signal-to-interference-plus-noise ratio (SINR) constraints at each communication receiver and the transmit power budget. Additionally, the work \cite{liu2020joint} proposed to jointly precode communication and radar waveforms to achieve maximum DoFs in waveform design. Recent progress on joint waveform design in MIMO-ISAC systems has been made in \cite{10680586} and \cite{10445319}, in which \cite{10680586} optimized a weighted combination of the sum rate and the CRB for target estimation and \cite{10445319} maximized the system energy efficiency by constraining the transmit power budget, communication SINR, and target estimation CRB.

The works mentioned previously accomplished a balance between S$\&$C to some extent, but they did not address the issue of range sidelobe control. As for now, there have been few studies on sidelobe control for ISAC systems. The work \cite{Liu2020} proposed a MIMO-ISAC waveform design framework that realized an integrated sidelobe level (ISL) reduction by minimizing a weighted sum of beampattern MSE, ISL, and MUI. Alternatively, the work \cite{liu2022joint} focused on maximizing the SINR at the radar output while ensuring communication performance, which also achieves the ISL reduction. However, the communication channels in the two works are assumed to be frequency-flat fading, whereas the ISAC signals capable of distinguishing symbol-level delays correspond to frequency-selective fading communication channels. %The work \cite{10771629} reduced the sidelobe level by minimizing the ISL under various S$\&$C constraints for MIMO-OFDM ISAC waveform.

In addition to the ISL metric, the peak sidelobe level (PSL) should be more important for sidelobe control. That's because the PSL at the radar receiver dictates the false alarm probability \cite{Sankuru2021}. High PSL can lead to false alarms \cite{haimovich2007mimo}, making PSL control significant for achieving a low false alarm rate (FAR). Despite its importance, the PSL minimization problem has received relatively little attention in both the radar signal \cite{Sankuru2021} and ISAC waveform design. One of the main difficulties in directly minimizing the PSL is that the design metric is not differentiable, and the corresponding optimization problem is a minimax problem. To the best of our knowledge, the existing works on PSL control focus on pure radar sensing, and PSL control in ISAC scenarios remains unexplored. For instance, the work \cite{Li2008} proposed a two-step scheme to control the PSL in MIMO radar. The work \cite{Song2016} approximated PSL suppression by minimizing an $\ell_{p}$ norm metric with large $p$ for single-antenna radar systems.

Taking the above factors into consideration, the goal of this paper is to design a downlink MIMO-ISAC waveform that maximizes the achievable rate at the communication receivers, makes the radar receiver's beampattern as close as possible to the desired one, and controls the PSL under the constraints of unimodular transmit sequences. However, as discussed in \cite{Li2008}, designing waveforms with good correlation properties under the constraint of unimodular sequences is already a complicated task, and it will be more challenging further to require good communication performance and the desired beampattern. To simplify the optimization problem and obtain more DoFs on waveform design, we consider the joint design of the receive filter and transmit sequence in this paper.
\subsection{Our Contributions}
The main contributions of this paper are as follows.
\begin{itemize}
  \item \textit{Unimodular ISAC waveform design with range sidelobe level control}: Many factors are taken into account in the problem formulation, including beampattern MSE, communication MUI, PSL, and unimodular transmit sequence. An optimization problem is formulated to minimize a weighted sum of the radar receive beampattern MSE and the MUI at the communication receiver. To avoid solving a minimax problem by minimizing the PSL directly, we control the PSL by constraining the level of sidelobe at each range bin. The formulated problem is a large-scale optimization problem with convex and nonconvex constraints.
  \item \textit{Efficient solution:} An inexact augmented Lagrangian method (ALM) algorithm is proposed to solve the formulated problem, where each ALM nonconvex subproblem is solved approximately to avoid the huge number of iterations for obtaining an exact stationary point. Specifically, a block successive upper-bound minimization (BSUM) scheme is custom-designed to solve the subproblems in the ALM algorithm, and the updates for all blocks of variable in the BSUM scheme admit closed-form solutions, which makes the proposed algorithm efficient.
  \item \textit{Convergence guarantee:} We analyze the convergence of the proposed inexact ALM algorithm with an adaptive penalty parameter. We show that the proposed algorithm is guaranteed to find a feasible stationary point of the formulated problem. This is the best that one can expect for this nonconvex optimization problem (with many nonconvex constraints).
\end{itemize}

Extensive simulation results are provided to demonstrate the effectiveness of the proposed algorithm. Specifically, Monte Carlo simulations are performed to evaluate the convergence performance of the proposed algorithm. The impacts of different system parameters on the system performance are examined. Finally, the proposed algorithm is compared with the ALM algorithm with a fixed penalty parameter and the modified work of \cite{Li2008} to show the superiority of the proposed algorithm.

It is worth mentioning that a similar inexact-ALM framework has been proposed in \cite{wu2024quantized}. In \cite{wu2024quantized}, radar sensing beamforming is achieved through quantized constant-envelope waveform design while satisfying the communication performance constraints. However, it does not address how to suppress the sidelobes of the correlation function. Low-range bin sidelobes and beampattern are equally important for good radar imaging. Furthermore, after considering sidelobe suppression, the convergence analysis of the optimization problem formulated in this paper becomes more complex compared to \cite{wu2024quantized} since the constraint terms introduced by the auxiliary variables in \cite{wu2024quantized} are linear, whereas those in this paper are nonlinear.

The rest of the paper is organized as follows. In Section \ref{sec:system_model}, we introduce the system model and formulate the optimization problem. In Section \ref{sec:prob_form_sol}, we propose an efficient algorithm for solving the formulated problem and analyze the convergence of the proposed algorithm. We present extensive simulation results in Section \ref{sec:num_res}. Finally, we conclude the paper in Section \ref{sec:conclusion}.

\textit{Notations:} We use $x$, $\mathbf{x}$, $\mathbf{X}$, and $\mathcal{X}$ to represent scalar, column vector, matrix, and set, respectively. The notation vec($\cdot$) represents the vectorization of a matrix by stacking its columns. The notations $\|\cdot\|_{1}$, $\|\cdot\|_{2}$, and $\|\cdot\|_{\text{F}}$ denote correspondingly the $\ell_1$, $\ell_2$, and Frobenius norms of a matrix, respectively. $\mathbb{C}$ and $\mathbb{R}$ denote the sets of complex and real numbers, respectively; $\mathcal{R}\{\cdot\}$ and $\mathcal{I}\{\cdot\}$ are the real and imaginary parts of a complex number, respectively. The superscript $(\cdot)^{*}$, $(\cdot)^{\text{T}}$ and $(\cdot)^{\text{H}}$ represent the conjugate, the transpose and the conjugate transpose operations, respectively. $\mathbb{I}_{\mathcal{A}}(A)$ denotes an indicator function of $\mathcal{A}$, and it takes $0$ if $A \in \mathcal{A}$ and $+\infty$, otherwise. $\otimes$ represents the Kronecker product. $\mathbf{A}_{[i:j, :]}$ represents the sub-matrix from the $i$-th row to the $j$-th row of a matrix $\mathbf{A}$, and $A[i,j]$ represents the element in the $i$-th row and $j$-th column of matrix $\mathbf{A}$. $\mathrm{tr}(\mathbf{A})$ means the trace of a matrix $\mathbf{A}$. $\mathbf{A} \succeq \mathbf{B}$ means $\mathbf{A} - \mathbf{B}$ is positive semidefinite. $[n]$ denotes $\{1, 2, \dots, n\}$ for a positive integer $n$.
\section{System Model and Problem Formulation}\label{sec:system_model}
\begin{figure*}[htpb]
  \centering
  \includegraphics[width=1.5\columnwidth]{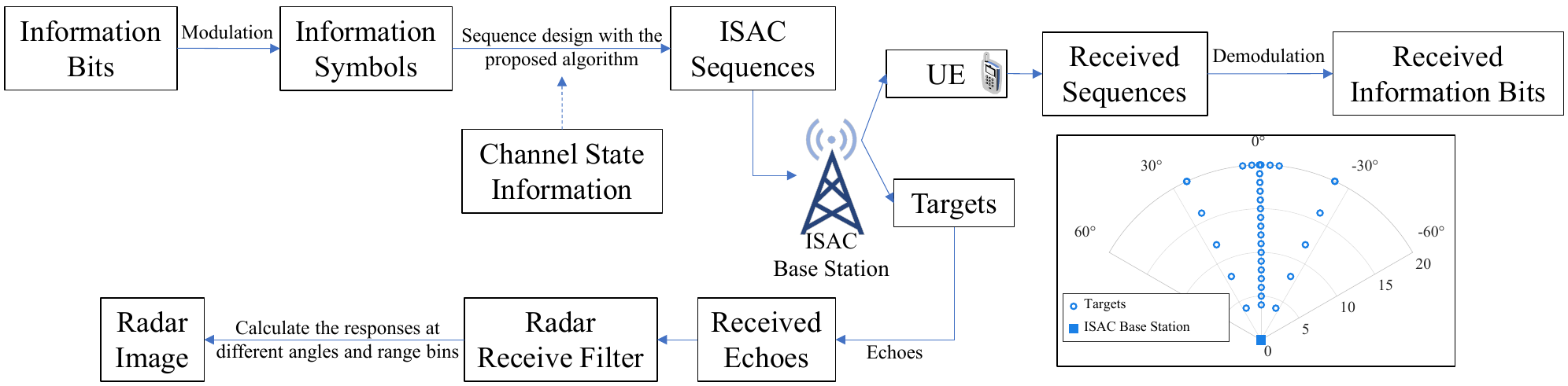}
  \caption{A dual-functional radar communication system.}
  \label{fig:ISAC_scenario}
\end{figure*}
As shown in Fig. \ref{fig:ISAC_scenario}, we consider an ISAC base station (BS) equipped with a uniform linear array (ULA) of $N_{\text{T}}$ antennas. $N_{\text{C}}$ single-antenna communication receivers are in the downlink transmission. We consider a colocated radar receive station with a $N_{\text{R}}$-antenna ULA. Below, we introduce the mathematical frameworks used in both S$\&$C functions and formulate the optimization problem.
\subsection{Communication Signal Model}
Considering the frequency-selective feature of the channel in the current communication system, \cite{3gpp.36.104}, we implement the channel model proposed in \cite{932706}, in which the channel between $m$-th receiver and $n$-th BS transmitting antenna, denoted as $\mathbf{H}_{m,n}$, is frequency-selective and assumed to be quasi-stationary during one block transmission. A $q$-ray model defines the discrete-time channel gains. The channel $\mathbf{H}_{m,n}$ can be characterized by a lower triangular Toeplitz matrix with the first column being $[h_0^{(m, n)}, h_{1}^{(m, n)}, \dots, h_{L}^{(m, n)}, \mathbf{0}_{T}]^{\text{T}}$, where $L = \lceil \frac{\tau_{\max}}{T_s} \rceil $, $T_s$ is the duration of one time-slot, $\tau_{\max}$ is the maximum delay spread, and $T$ is the length of the block.

Due to the multi-taps feature of the channel, inter-block interference (IBI) occurs between every two consecutive transmissions (blocks). To eliminate IBI, a cyclic prefix (CP) of length $L$ is added at each transmission and will be removed at the receiver. The noiseless received signal at the $m$-th receiver is given by $\mathbf{y}_{m} = \sum_{n=1}^{N_{\text{T}}} \tilde{\mathbf{H}}_{m,n} \mathbf{x}_{n}$, where $\tilde{\mathbf{H}}_{m,n} = \bm{\Pi}_{\text{cp}} \mathbf{H}_{m,n} \bm{\Gamma}_{\text{cp}}$ is a $T \times T$ matrix, $\bm{\Gamma}_{\text{cp}} = [\mathbf{I}_{\text{cp}}, \mathbf{I}_{T}]^{\text{T}}$ and $\bm{\Pi}_{\text{cp}} = [\mathbf{0}_{T \times L}, \mathbf{I}_{T}]$ denote the CP-inducing and CP-removing matrices, respectively, and $\mathbf{I}_{\text{cp}}$ contains the last $L$ columns of a $T$-dimensional identity matrix $\mathbf{I}_{T}$. Then, the received signals at all $N_{\text{C}}$ users can be represented as
\begin{equation}\label{eq:Comm_model}
  \mathbf{y} = \mathbf{H} \mathbf{x} + \mathbf{n},
\end{equation}
where $\mathbf{y} = [\mathbf{y}_{1}^{\text{T}}, \mathbf{y}_{2}^{\text{T}}, \dots, \mathbf{y}_{N_{\text{C}}}^{\text{T}}]^{\text{T}}$ with $\mathbf{y}_{i} \in \mathbb{C}^{T \times 1}$ being the received signal at the $i$-th user, $\mathbf{x} = [\mathbf{x}_{1}^{\text{T}}, \mathbf{x}_{2}^{\text{T}}, \dots, \mathbf{x}_{N_{\text{T}}}^{\text{T}}]$ with $\mathbf{x}_{i} \in \mathbb{C}^{T \times 1}$ being the $i$-th row vector of the transmission signal matrix $\mathbf{X} \in \mathbb{C}^{N_T \times T}$, $\mathbf{n} \in \mathbb{C}^{N_{\text{T}}(T+L) \times 1}$ is an additive white Gaussian noise (AWGN) vector with power $\sigma_{n}^2$, and $\mathbf{H} \in \mathbb{C}^{N_{\text{C}}T \times N_{\text{T}} T}$ is defined as
\begin{equation}
  \mathbf{H} = \left[ \begin{smallmatrix}
      \tilde{\mathbf{H}}_{1,1} & \tilde{\mathbf{H}}_{1,2} & \dots & \tilde{\mathbf{H}}_{1,N_{\text{T}}}\\
      \tilde{\mathbf{H}}_{2,1} & \tilde{\mathbf{H}}_{2,2} & \dots & \tilde{\mathbf{H}}_{2,N_{\text{T}}}\\
      \vdots & & \ddots & \vdots \\
      \tilde{\mathbf{H}}_{N_{\text{C}},1} & \tilde{\mathbf{H}}_{N_{\text{C}},2} & \dots & \tilde{\mathbf{H}}_{N_{\text{C}},N_{\text{T}}}
    \end{smallmatrix} \right].
\end{equation}
We assume that the BS has perfect channel state information (CSI), which allows us to clearly highlight the performance gain and design principles of the proposed system under ideal conditions. The waveform design under partial or statistical CSI, as in \cite{yao2025exploring}, will be our future work.

Let $\mathbf{S} \in \mathbb{C}^{N_{\text{C}} \times T}$ denote the transmitted information symbol matrix, where each entry of $\mathbf{S} $ is randomly drawn from a given constellation. The received signal at the communication users can then be represented in vector form as
\begin{equation}
  \mathbf{y} = \mathbf{s} + (\mathbf{H} \mathbf{x} - \mathbf{s}) + \mathbf{n},
\end{equation}
where $\mathbf{s} = [\mathbf{s}_{1}^{\text{T}}, \mathbf{s}_{2}^{\text{T}}, \dots, \mathbf{s}_{N_{\text{C}}}^{\text{T}}]^{\text{T}}$ with $\mathbf{s}_{i} \in \mathbb{C}^{T \times 1}$ being the $i$-th row vector of $\mathbf{S}$, the term $\mathbf{H} \mathbf{x} - \mathbf{s}$ can be viewed as the MUI\cite{ISAC1} that interferes with the symbol detection at the communication receiver side. The received SINR at the $i$-th user is defined as
\begin{equation}
  \gamma_i = \frac{\mathbb{E}_{\mathbf{s}_{i}}\{\|\mathbf{s}_{i}\|_{2}^{2}\}}{\mathbb{E}_{\mathbf{s}}}\{\|\mathbf{H}_{i} \mathbf{x} - \mathbf{s}\|_{2}^{2}\} + \sigma_n^2,
\end{equation}
where $\mathbf{H}_{i} = [\tilde{\mathbf{H}}_{i,1}, \tilde{\mathbf{H}}_{i,2}, \dots, \tilde{\mathbf{H}}_{i,N_{\text{T}}}]$. The achievable downlink sum rate of the users can be given as
\begin{equation}
  R = \sum_{i=1}^{N_{\text{C}}} \log_{2}(1 + \gamma_i).
\end{equation}

Given that $\mathbb{E}_{\mathbf{s}_{i}}\{\|\mathbf{s}_{i}\|_{2}^{2}\}$ is a fixed value for a specific constellation strategy, minimizing $P_{\text{MUI}} = \|\mathbf{H}\mathbf{x} - \mathbf{s}\|_{2}^{2}$ leads to an increase in SINR, thereby indicating a higher achievable sum rate\footnote{Minimizing $P_{\text{MUI}}$ is not equivalent to maximizing sum rate. We use this metric to simplify the optimization problem.}.

\subsection{Signal Model for Radar Imaging}
\begin{figure}[t]
  \centering
  \includegraphics[width=0.8\columnwidth]{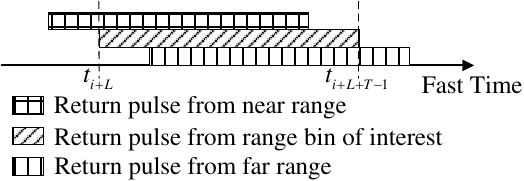}
  \caption{The overlapped returned pulses.}
  \label{fig:radar_iamging_signal_model}
\end{figure}

As shown in Fig. \ref{fig:radar_iamging_signal_model}, after removing the CP, the received echos for radar imaging at the $i$-th range bin, $\mathbf{D}_{(i)} \in \mathbb{C}^{N_{R} \times T}$, are obtained from the $t_{i+L}$-th time slot to the $t_{i+L+T-1}$-th time slot, where $t_{i+L}$ is the starting time slot of the $i$-th range bin, and $\mathbf{D}_{(i)} \in \mathbb{C}^{N_{R} \times T}$ can be expressed as \cite[eq. (13)]{Li2008}
\begin{multline}\label{eq:radar_echo}
  \mathbf{D}_{(i)} = \sum_{p_{(i)} = 1}^{P_{(i)}} h_{p_{(i)}} \mathbf{a}(\theta_{p_{(i)}}) \mathbf{v}(\theta_{p_{(i)}})^{\text{T}} \mathbf{X} \\
  + \sum_{k_{(i)} \in \mathbf{\Omega}_{K}} \sum_{p_{k_{(i)}}^{\prime} = 1}^{P_{k_{(i)}}^{\prime}} h_{p_{k_{(i)}}^{\prime}} \mathbf{a}(\theta_{p_{k_{(i)}}^{\prime}}) \mathbf{v}(\theta_{p_{k_{(i)}}^{\prime}})^{\text{T}} \mathbf{X} \mathbf{J}_{k} + \mathbf{Z}_{\text{R}},
\end{multline}
where $\mathbf{X}$ is the transmitted ISAC waveform defined above, $\mathbf{Z}_{\text{R}}$ is the additive Gaussian noise matrix with zero mean and covariance matrix $\mathbf{R}_{\text{s}}$, $\{\mathbf{a}(\theta_{p_{(i)}}) \in \mathbb{C}^{N_{R} \times 1}\}_{p_{(i)}=1}^{P_{(i)}}$ with its $n$-th element being $e^{-j 2 \pi n \frac{d}{\lambda} \sin(\theta_{p, i})}$, $n=0, 1, \dots, N_R-1$, and $\{\mathbf{v}(\theta_{p_{(i)}}) = [1, e^{-j 2 \pi \frac{d}{\lambda} \sin(\theta_{p, i})}, \dots, e^{-j 2 \pi (N_{T}-1) \frac{d}{\lambda} \sin(\theta_{p, i})}] \in \mathbb{C}^{N_{T} \times 1}\}_{p_{(i)} = 1}^{P_{(i)}}$ are the radar receive and transmit steering vectors for the $P_{(i)}$ targets in the $i$-th range bin, respectively, $d = \lambda/2$, and $\{h_{p_{(i)}} \in \mathbb{C}\}_{p_{(i)}=1}^{P_{(i)}}$ are complex amplitudes proportional to the RCS of these $P_{(i)}$ targets. The parameters with subscript $p_{k_{(i)}}^{\prime}$ are defined for the $P_{k_{(i)}}^{\prime}$ scatterers at the $k_{(i)}$-th range bin with the same meanings, the temporal shifting matrix $\mathbf{J}_{k}$ is defined as
$$\mathbf{J}_{k} = \mathbf{J}_{-k}^{\text{T}} = \left[\begin{smallmatrix}
      \mathbf{0}_{(T - k) \times k} & \mathbf{I}_{T - k}  \\
      \mathbf{I}_{k}   & \mathbf{0}_{k\times (T - k)} \\
    \end{smallmatrix}\right],$$
and $\mathbf{\Omega}_{K} = \{-K, \dots, -1, 1, 2, \dots, K\}$, where $K$ is the maximum difference of arrival times between backscattered signals from the range bin of interest and signals from neighboring range bins.

To obtain more DoFs on the waveform design, we jointly design the radar receive filter and transmit sequence. Denote the receive filter at the radar receiver as $\mathbf{F} \in \mathbb{C}^{N_{\text{R}} \times T}$, the radar image at angle $\theta$ and range bin $i$ is given by
\begin{equation}\label{eq:DF_radar}
 r_{\theta, i} = |\mathbf{a}(\theta)^{\text{H}} \mathbf{D}_{(i)} \mathbf{F}^{\text{H}} \mathbf{a}(\theta)|.
\end{equation}
Under the signal models \eqref{eq:radar_echo} and \eqref{eq:DF_radar}, we consider optimizing two metrics to improve radar imaging performance. One is the radar receive beampattern MSE, defined as $\| \mathbf{X} \mathbf{F}^{\text{H}} - \mathbf{R}_{\text{d}} \|_{\text{F}}$, measuring how close the receive beampattern approximates the desired one, where $\mathbf{R}_{\text{d}}$ is the desired spatial response determined by the prior knowledge about the targets \cite{wu2024quantized} so that we can have stronger angle responses of the targets.For example, if we roughly know there are three targets at $-10^{\circ}$, $0^{\circ}$, and $10^{\circ}$, we can solve the correlation matrix $\mathbf{R}_{\text{d}}$ through the algorithm in \cite{varadhan2008simple}. Otherwise, we set $\mathbf{R}_{\text{d}}$ as an identity matrix, which corresponds to an omnidirectional beampattern, if we have no knowledge about the targets. Another metric is the maximum sidelobe level $\max_{k \in \mathbf{\Omega}_{K}}\{\|\mathbf{X} \mathbf{J}_{k} \mathbf{F}^{\text{H}}\|_{\text{F}}\}$, which is required to be less than a preset level $\xi$ for all $k \in \mathbf{\Omega}_{K}$ to reduce the clutters from neighboring range bins and to ensure a clear response on the range bin of interest.
\subsection{Problem Formulation}
Based on the above discussions, the joint design problem of the radar receive filter and the ISAC waveform is formulated to minimize a weighted sum of the receive radar beampattern MSE and MUI at communication receivers under the constraints of unimodular transmit sequence and the full power limit of the radar receive filter\footnote{We adopt the full-power radar receive filter to ensure that the mainlobe energy remains as high as possible while suppressing the sidelobe level. Although this can be achieved by maximizing the sidelobe-to-mainlobe ratio, it leads to a fractional programming problem, increasing the complexity of the solution. In future work, we will consider using the sidelobe-to-mainlobe ratio as an optimization objective or constraint.}. The optimization problem is presented below:
\begin{subequations}
  \begin{align}
     & \min_{\mathbf{F}, \mathbf{X}} f(\mathbf{F}, \mathbf{X})     \label{prob:prob_origin_obj}                                                            \\
     & \;\text{s.t. } \| \mathbf{F} \|_{\text{F}}^2 = P_{\text{F}},    \label{prob:PSL_row_b}                                                              \\
     & \;\;\;\;\;\;\, \| \mathbf{X} \mathbf{J}_{k} \mathbf{F}^{\text{H}} \|_{\text{F}} \leq \xi,~\forall~k \in \mathbf{\Omega}_{K}, \label{prob:PSL_row_c} \\
     & \;\;\;\;\;\;\; |x_{ij}| = P_{x}, \forall~x_{ij} \in \mathbf{X}\label{prob:PSL_row_d},
  \end{align}\label{prob:prob_origin}
\end{subequations}
where  $f(\mathbf{F}, \mathbf{X}) = \alpha \| \mathbf{X} \mathbf{F}^{\text{H}} - \mathbf{R}_{\text{d}} \|_{\text{F}}^2 + (1 - \alpha) \|\mathbf{H} \mathbf{x} - \mathbf{s}\|_{2}^2$, and \eqref{prob:PSL_row_d} refers to the unimodular constraint on the transmit sequence.

Solving problem \eqref{prob:prob_origin} is not straightforward due to the presence of both convex PSL constraints and nonconvex constraints on the receive filter and transmit sequence. In the next section, we propose an inexact ALM algorithm for solving problem \eqref{prob:prob_origin} efficiently. ``Inexact'' here refers to solving the subproblems in the ALM algorithm inexactly. By doing so, we can significantly reduce the computational cost of solving the ALM subproblems without sacrificing the solution quality.
\section{Proposed Approach}\label{sec:prob_form_sol}
In this section, we propose an inexact ALM algorithm for solving problem \eqref{prob:prob_origin}. More specifically, we first introduce the main idea of the proposed inexact ALM algorithm in Section \ref{sec:main_idea_of_ALM}. Then, we propose a scheme for finding a feasible point of problem \eqref{prob:prob_origin} in Section \ref{sec:feasible_pt_details}, which plays an important role in guaranteeing the convergence of the proposed algorithm to a \emph{feasible} stationary point. We custom-design a BSUM algorithm for efficiently solving the ALM subproblem in Section \ref{sec:BSUM_details}. Finally, we analyze the convergence of the proposed algorithm in Section \ref{sec:convergence_analysis}.
\subsection{Framework of Proposed ALM Algorithm}\label{sec:main_idea_of_ALM}
The inexact ALM algorithm is proposed to solve problem \eqref{prob:prob_origin}. Its basic idea is to decompose the original optimization problem into smaller and more manageable subproblems. At each iteration, the algorithm updates primal variables associated with each ALM subproblem with a penalty term for penalizing the violation of the constraints in the original problem, followed by an update of the dual variables. Unlike the classic ALM algorithm, which solves each subproblem exactly, we solve each subproblem \textit{inexactly} at each iteration for computational efficiency.

To present the inexact ALM algorithm, we first reformulate problem \eqref{prob:prob_origin} by introducing auxiliary variables $\bm{C}= (\mathbf{C}_{k})_{k \in \mathbf{\Omega}_{K}}$ with $\mathbf{C}_{k} \in \mathbb{C}^{N_{T} \times N_{R}}$ for each $k \in \mathbf{\Omega}_{K}$. The problem \eqref{prob:prob_origin} becomes
\begin{equation}
  \begin{aligned}\label{prob:FX_with_C}
     & \min_{\mathbf{F}, \mathbf{X}, \bm{C}} f(\mathbf{F}, \mathbf{X})                             \\
     & \;\; \text{s.t. } \| \mathbf{F} \|_{\text{F}}^2 = P_{\text{F}},                             \\
     & \;\;\;\;\;\;\;\, \mathbf{X} \mathbf{J}_{k} \mathbf{F}^{\text{H}} = \mathbf{C}_{k},          \\
     & \;\;\;\;\;\;\;\, \|\mathbf{C}_{k}\|_{\text{F}} \leq \xi,~\forall~k \in \mathbf{\Omega}_{k}, \\
     & \;\;\;\;\;\;\;\; |x_{ij}| = P_{x}, \forall~x_{ij} \in \mathbf{X}.
  \end{aligned}
\end{equation}
By denoting the Lagrange multipliers associated with the constraints $\mathbf{X} \mathbf{J}_{k} \mathbf{F}^{\text{H}} = \mathbf{C}_{k}$ for all $k \in \mathbf{\Omega}_K$ as $\bm{U} = (\mathbf{U}_{k})_{k \in \mathbf{\Omega}_{K}}$, the corresponding augmented Lagrangian function (ALF) \cite{bertsekas2016nonlinear} of problem \eqref{prob:FX_with_C} is
\begin{align}\label{eq:ALF_origin}
   & \mathcal{L}_{\rho}(\bm{C}, \mathbf{F}, \mathbf{X}; \bm{U}) =  f(\mathbf{F}, \mathbf{X}) + \sum_{k \in \mathbf{\Omega}_{K}} \Big[\frac{\rho}{2} \|\mathbf{X} \mathbf{J}_{k} \mathbf{F}^{\text{H}} - \mathbf{C}_{k}\|_{\text{F}}^2 \nonumber \\
   & \qquad \quad \qquad \qquad + \mathcal{R}\{tr[{\mathbf{U}_{k}}^{\text{H}} (\mathbf{X} \mathbf{J}_{k} \mathbf{F}^{\text{H}} - \mathbf{C}_{k})]\}\Big].
\end{align}

The proposed ALM algorithm is based on the above ALF and mainly consists of four main components, which are the initialization, the solution of the ALM subproblem, the update of the Lagrange variable (also called dual variable), and the update of the penalty parameter. In the next, we introduce them one by one and highlight their important roles in the whole ALM algorithm. To simplify the notations, the point $(\bm{C}, \mathbf{F}, \mathbf{X})$ is represented by a multiplet $\bm{z} = (\bm{C}, \mathbf{F}, \mathbf{X})$.
\subsubsection{Initialization}To initialize the proposed algorithm, we choose a feasible point $\bm{z}^{(\text{feas})} \in \mathcal{S} \cap \mathcal{S}_{0}$, where
$$
  \begin{aligned}
    \mathcal{S}_{0} & = \left\{(\bm{C}, \mathbf{F}, \mathbf{X}) \mid \mathbf{X} \mathbf{J}_{k} \mathbf{F}^{\text{H}} = \mathbf{C}_{k},~\forall~k \in \mathbf{\Omega}_{k}\right\},                                             \\
    \mathcal{S}     & = \{(\bm{C}, \mathbf{F}, \mathbf{X}) \mid \mathbf{C}_{k} \in \mathcal{S}_{\text{C}},~\forall~k \in \mathbf{\Omega}_{k}; \mathbf{F} \in \mathcal{S}_{\text{F}}; \mathbf{X} \in \mathcal{S}_{\text{X}}\},
  \end{aligned}
$$
and
\begin{equation}
  \begin{aligned}
    \mathcal{S}_{\text{C}} & = \left\{\mathbf{C} \in \mathbb{C}^{N_{T} \times N_{T}} \mid \|\mathbf{C}\|_{\text{F}} \leq \xi\right\}, \\
    \mathcal{S}_{\text{F}} & = \{\mathbf{F} \in \mathbb{C}^{N_{T} \times T} \mid \|\mathbf{F}\|_{\text{F}}^2 = P_{\text{F}}\},        \\
    \mathcal{S}_{\text{X}} & = \{\mathbf{X} \in \mathbb{C}^{N_{T} \times T} \mid  |x_{ij}| = P_{x}, \forall~x_{ij} \in \mathbf{X}\}.
  \end{aligned}
\end{equation}
The details of finding a feasible point of problem \eqref{prob:FX_with_C} is relegated to Section \ref{sec:feasible_pt_details} to maintain smoothness in presenting the proposed inexact ALM algorithm.

By choosing the penalty parameter $\rho^{(0)} > 0$, choosing an arbitrary initial point $\bm{z}^{(0)} \in \mathcal{S}$, and setting a finite Lagrange multiplier $\bm{U}^{(0)}$, we can choose a finite constant number $\zeta$ satisfying
\begin{equation}\label{eq:zeta_the_upper_bound}
  \zeta \geq \max\left\{f(\mathbf{F}^{(\text{feas})}, \mathbf{X}^{(\text{feas})}), \mathcal{L}_{\rho^{(0)}}(\bm{z}^{(0)}; \bm{U}^{(0)})\right\}.
\end{equation}
The upperboundness of the objective function and the ALF are important in proving the convergence of the proposed ALM algorithm to a feasible stationary point. In our proposed ALM algorithm, the penalty parameter is adaptively updated. The adaptive update of the penalty parameter enables a fast convergence but also makes the convergence analysis difficult. The upperboundness here can guarantee that any limit point of the sequence generated by the proposed algorithm is always a feasible stationary point, even when the penalty parameter becomes positive infinity. More details on the convergence analysis will be presented in Section \ref{sec:convergence_analysis}.
\subsubsection{Solving the ALM Subproblem Inexactly}
At the $\ell$-th iteration, the classic ALM algorithm updates the primal variables by solving the following ALM subproblem:
\begin{equation}\label{prob:sub_problem_ALM}
  \bm{z}^{(\ell+1)} \in \argmin_{\substack{\bm{z} \in \mathcal{S}}} \mathcal{L}_{\rho^{(\ell)}}( \bm{z}; \bm{U}^{(\ell)}).
\end{equation}
However, problem \eqref{prob:sub_problem_ALM} is nonconvex due to the nonconvex constraints of full power radar receiving filter and unimodular transmit sequence. To improve the computational efficiency, we propose to inexactly solve problem \eqref{prob:sub_problem_ALM} to an $\varepsilon^{(\ell)}$-stationary point, where the $\varepsilon$-stationary point is defined below:
\begin{definition}[$\varepsilon$-stationary point of problem \eqref{prob:sub_problem_ALM}]\label{def_1}
  For a fixed point $\bar{\bm{U}}$, the point $\bar{\bm{z}} \in \mathcal{S}$ is called an $\varepsilon$-stationary point of problem \eqref{prob:sub_problem_ALM} if there exist $\bar{\bm{A}} \in \partial \mathbb{I}_{\mathcal{S}_{\rm F}}(\bar{\mathbf{F}})$, $\bar{\bm{B}} \in \partial \mathbb{I}_{\mathcal{S}_{\rm X}}(\bar{\mathbf{X}})$, and $\bar{\bm{D}}_{k} \in \partial \mathbb{I}_{\mathcal{S}_{\rm C}}(\bar{\mathbf{C}}_{k})$ for all $k \in \mathbf{\Omega}_{K}$ such that
  \begin{equation}
    \begin{aligned}
      \|\bar{\bm{A}} + \nabla_{\mathbf{F}} \mathcal{L}_{\rho}(\bar{\bm{z}}; \bar{\bm{U}})\|_{{\rm F}}         & \leq \varepsilon,                                    \\
      \|\bar{\bm{B}} + \nabla_{\mathbf{X}} \mathcal{L}_{\rho}(\bar{\bm{z}}; \bar{\bm{U}})\|_{{\rm F}}         & \leq \varepsilon,                                    \\
      \|\bar{\bm{D}}_{k} + \nabla_{\mathbf{C}_{k}} \mathcal{L}_{\rho}(\bar{\bm{z}}; \bar{\bm{U}})\|_{{\rm F}} & \leq \varepsilon,~\forall~k \in \mathbf{\Omega}_{K}.
    \end{aligned}\nonumber
  \end{equation}
\end{definition}
\noindent The notations $\partial \mathbb{I}_{\mathcal{S}_{\text{F}}}(\mathbf{F})$, $\partial \mathbb{I}_{\mathcal{S}_{\text{X}}}(\mathbf{X})$, and $\partial \mathbb{I}_{\mathcal{S}_{\text{C}}}(\mathbf{C}_{k})$ are subdifferentials of indicator functions; see \cite{Hallak2023} and \cite[Corollary 8.20]{drusvyatskiy2020convex} for more details. The $\varepsilon$-stationary point defined in Definition 1 reduces to the standard stationary point when setting $\varepsilon=0$.

In inexactly solving the ALM subproblem in \eqref{prob:sub_problem_ALM}, we also require that the inexact solution $z^{(\ell+1)}$ also satisfies that the value of the ALF is bounded by
\begin{equation}\label{ineq:upper_bound_requirement}
  \mathcal{L}_{\rho^{(\ell)}}( \bm{z}^{(\ell+1)}; \bm{U}^{(\ell)}) \leq \zeta,
\end{equation}
where $\zeta$ is defined in \eqref{eq:zeta_the_upper_bound}. To find an $\varepsilon$-stationary point that satisfies (9), we shall develop a BSUM algorithm, which will be elaborated in detail in Section \ref{sec:BSUM_details}. This algorithm offers closed-form updates for all blocks of variable and is guaranteed to find an $\varepsilon$-stationary point within a finite number of iterations.

\subsubsection{Updating Lagrange Multipliers}
Once the $\varepsilon^{(\ell)}$-stationary point is obtained, we update the Lagrange multipliers in $\bm{U}$ for all $k \in \mathbf{\Omega}_{K}$ according to the following updating rule:
\begin{subequations}
  \begin{align}
     & \tilde{\mathbf{U}}_{k} = \mathbf{U}_{k}^{(\ell)} + \rho^{(\ell)} (\mathbf{X}^{(\ell+1)} \mathbf{J}_{k} {\mathbf{F}^{(\ell+1)}}^{\text{H}} - \mathbf{C}_{k}^{(\ell+1)}), \label{eq:uk_update} \\
     & U_{k}^{(\ell+1)}[i,j] = \begin{cases}
                                 \frac{u_{\max}}{|\tilde{U}[i, j]|} \tilde{U}[i, j], & \text{if } |\tilde{U}[i, j]| > u_{\max}; \\
                                 \tilde{U}[i, j],                                    & \text{otherwise}.
                               \end{cases}\label{eq:uk_update_bound}
  \end{align}
\end{subequations}
In the above, $\mathbf{U}_{k}$ is first updated through the standard updating rule \cite{bertsekas2016nonlinear} in \eqref{eq:uk_update}; then, each element in $\mathbf{U}_{k}$ is projected onto an interval $[-u_{\max}, u_{\max}]$ with $u_{\max}>0$ being a preset constant. The projection onto the bounded set in \eqref{eq:uk_update} guarantees that the Lagrange multipliers are uniformly bounded and play a central role in guaranteeing the convergence of the proposed algorithm. The boundness of the multiplier sequence in the nonconvex setting remains an open research question \cite{Hallak2023}. Readers can refer to \cite{Hallak2023} for detailed discussions.

\subsubsection{Updating the Penalty Parameter}
We propose an adaptive update rule for the penalty parameter to accelerate the convergence speed. Denote the violation of the constraints after the $\ell$-th iteration as $v^{(\ell+1)}$, i.e.,
\begin{equation}\label{eq:constraints_violation}
  v^{(\ell+1)} = \sqrt{\sum_{k \in \mathbf{\Omega}_{K}} \| \mathbf{X}^{(\ell+1)} \mathbf{J}_{k} {{\mathbf{F}}^{(\ell+1)}}^{\text{H}} - \mathbf{C}_{k}^{(\ell+1)}\|_{\text{F}}^2}.
\end{equation}
The penalty parameter $\rho^{(\ell+1)}$ will be updated through the following rule:
\begin{equation}\label{eq:rho_update}
  \rho^{(\ell+1)} = \begin{cases}
    \gamma \rho^{(\ell)}, & \text{if } v^{(\ell+1)} > \delta v^{(\ell)}; \\
    \rho^{(\ell)},        & \text{otherwise}.
  \end{cases}
\end{equation}
where $\delta > 0$ and $\gamma > 1$. The update rule in \eqref{eq:rho_update} will increase the penalty parameter if the violation at the current iteration is not reduced sufficiently compared with the previous one. By dynamically adjusting the penalty parameter based on the current state of constraint violations and dual variable updates, this approach helps maintain an effective balance between enforcing constraint feasibility and minimizing the objective function.

\begin{algorithm}[t]
  \caption{An Inexact ALM Algorithm for Problem \eqref{prob:prob_origin}}
  \begin{algorithmic}[1]\label{alg:RALM}
    \REQUIRE Initial point $\bm{z}^{(0)}$, $\bm{U}^{(0)}$, $\{\varepsilon^{(\ell)}\}_{\ell \geq 0}$ strictly decreasing with $\lim_{\ell \rightarrow \infty} \varepsilon^{(\ell)} = 0$, penalty parameter $\rho^{(0)}$.
    \REPEAT
    \STATE Obtain an $\varepsilon^{(\ell)}$-stationary point $\bm{z}^{(\ell+1)}$ by Algorithm \ref{alg:inexact_ALM_sol} (details provided in Section \ref{sec:BSUM_details});
    \STATE Update the dual variables $\bm{U}^{(\ell+1)}$ by \eqref{eq:uk_update};
    \STATE Update the penalty parameter $\rho^{(\ell+1)}$ by \eqref{eq:rho_update};
    \STATE $\ell \gets \ell + 1$;
    \UNTIL{certain stopping criteria;}
    \ENSURE $\bm{z}^{(\ell+1)}$.
  \end{algorithmic}
\end{algorithm}

The proposed inexact ALM algorithm for solving problem \eqref{prob:FX_with_C} is summarized in Algorithm \ref{alg:RALM}. By setting a positive sequence $\{\varepsilon^{(\ell)}\}_{\ell \geq 0}$ with $\varepsilon^{(\ell)} \rightarrow 0$ as $\ell \rightarrow +\infty$, the limit point of the sequence generated by Algorithm \ref{alg:RALM}, $(\bm{z}^{(\infty)}, \bm{U}^{(\infty)})$, is a stationary point of the ALF in \eqref{eq:ALF_origin}. Besides, since $\mathbf{X}^{(\infty)} \mathbf{J}_{k} {{\mathbf{F}}^{(\infty)}}^{\text{H}} = \mathbf{C}_{k}^{(\infty)}$ holds for all $k \in \mathbf{\Omega}_{K}$, which will be proved further ahead, then $\bm{z}^{(\infty)}$ is a feasible stationary point of problem \eqref{prob:FX_with_C}. According to \cite{Bolte2018}, any feasible stationary point of \eqref{prob:FX_with_C} is also a feasible stationary point of \eqref{prob:prob_origin}, i.e., $(\mathbf{F}^{(\infty)}, \mathbf{X}^{(\infty)})$ is a feasible stationary point of \eqref{prob:prob_origin}. The detailed convergence analysis will be given in Section \ref{sec:convergence_analysis}.
\subsection{A Feasible Point of Problem \eqref{prob:FX_with_C}}\label{sec:feasible_pt_details}
In this subsection, we focus on finding a feasible point of problem \eqref{prob:FX_with_C}, i.e., a pair of matrices $(\mathbf{F}, \mathbf{X}) \in \mathcal{S}_{\text{F}} \times \mathcal{S}_{\text{X}}$ such that
\begin{equation}\label{ineq:feasible_condition}
  \| \mathbf{X} \mathbf{J}_{k} \mathbf{F}^{\text{H}} \|_{\text{F}} \leq \xi,~\forall~k \in \mathbf{\Omega}_{K}.
\end{equation}
The auxiliary variable $\bm{C}$ can be easily obtained by setting $\mathbf{C}_{k} = \mathbf{X} \mathbf{J}_{k} \mathbf{F}^{\text{H}}$ for all $k \in \mathbf{\Omega}_{K}$.

We formulate two subproblems for finding $(\mathbf{F}, \mathbf{X}) \in \mathcal{S}_{\text{F}} \times \mathcal{S}_{\text{X}}$ that satisfies \eqref{ineq:feasible_condition}. To be specific, after randomly initializing $(\mathbf{F}, \mathbf{X})$, the first subproblem is formulated to reduce the ISL, i.e., $\sum_{k \in \mathbf{\Omega}_{K}} \| \mathbf{X} \mathbf{J}_{k} \mathbf{F}^{\text{H}} \|_{\text{F}}^2$, concerning $\mathbf{X}$. We update $\mathbf{X}$ several times to minimize the ISL: if $\max_{k \in \mathbf{\Omega}_{K}} \{\| \mathbf{X} \mathbf{J}_{k} \mathbf{F}^{\text{H}} \|_{\text{F}}\}$ is already less than $\xi$, we then find a feasible point; otherwise, we formulate the second subproblem to minimize the ISL concerning $\tilde{\mathbf{F}} = \mathbf{F}^{\text{H}} \mathbf{F}$. When solving for the variable $\mathbf{X}$, the following lemma is useful.
\begin{lemmaNoParens}[\cite{wang2016design}]\label{lm:MM}
  Given Hermitian $\mathbf{M} \in \mathbb{C}^{n \times n}$ and $\mathbf{Z} \in \mathbb{C}^{m \times m}$ and any $\mathbf{X}^{(t)} \in \mathbb{C}^{m \times n}$, the function $\mathrm{tr}(\mathbf{Z} \mathbf{X} \mathbf{M} \mathbf{X}^{\rm H})$ can be majorized by
  $$\lambda \| \mathbf{X} \|_{\rm F}^2 + 2\mathcal{R} \left\{ tr\left[ (\mathbf{Z} \mathbf{X}^{(t)} \mathbf{M} - \lambda \mathbf{X}^{(t)})^{\rm H} \mathbf{X}\right] \right\} + C,$$
  where $C = tr[(\mathbf{X}^{(t)})^{\rm H} (\lambda \mathbf{X}^{(t)} - \mathbf{Z} \mathbf{X}^{(t)} \mathbf{M})]$ is irrelevant to $\mathbf{X}$ and $\lambda$ satisfies $\lambda \mathbf{I} \succeq \mathbf{M}^{\rm T} \otimes \mathbf{Z}$.
\end{lemmaNoParens}
\subsubsection{Reducing the ISL Regarding $\mathbf{X}$}
By choosing any $\mathbf{F} \in \mathcal{S}_{\text{F}}$, and denoting $\mathbf{\Theta} = \sum_{k \in \mathbf{\Omega}_{K}} \mathbf{J}_{k} \mathbf{F}^{\text{H}} \mathbf{F} \mathbf{J}_{k}^{\text{H}}$, the problem of reducing the ISL regarding $\mathbf{X}$ is formulated as
\begin{equation}\label{prob:X_feasible}
  \begin{aligned}
     & \min_{\mathbf{X}} \mathrm{tr}(\mathbf{X} \mathbf{\Theta} \mathbf{X}^{\text{H}}) \\
     & \;\text{s.t. } \mathbf{X} \in \mathcal{S}_{\text{X}}.
  \end{aligned}
\end{equation}

According to \textit{Lemma \ref{lm:MM}}, the tightest upper bound $\lambda$ to majorize $\mathrm{tr}(\mathbf{X} \mathbf{\Theta} \mathbf{X}^{\text{H}})$ is the maximum eigenvalue of $\mathbf{\Theta}$, denoted by $\lambda_{\max}(\mathbf{\Theta})$. However, $\mathbf{\Theta}$ may change after updating $\mathbf{F}$. To avoid the high computational cost of computing $\lambda_{\max}(\mathbf{\Theta})$, an alternative upper bound $\lambda_{\theta}$ is implemented, where
\begin{equation}
  \lambda_{\theta} = \|\mathbf{\Theta}\|_{1} \geq \lambda_{\max}(\mathbf{\Theta}). \nonumber
\end{equation}
Then, the quadratic term $\mathrm{tr}(\mathbf{X} \mathbf{\Theta} \mathbf{X}^{\text{H}})$ can be majorized as follows:
\begin{multline}
  \mathrm{tr}(\mathbf{X} \mathbf{\Theta} \mathbf{X}^{\text{H}}) \leq \lambda_{\theta} \|\mathbf{X}\|_{\text{F}}^2  \\
  + 2 \mathcal{R}\left\{ tr\left[\left(\mathbf{X}^{(t)} \mathbf{\Theta} - \lambda_{\theta} \mathbf{X}^{(t)}\right)^{\text{H}} \mathbf{X} \right] \right\} + C_{\text{X}},
\end{multline}
where $C_{\text{X}}$ is the term irrelevant to $\mathbf{X}$ and $\mathbf{X}^{(t)}$ is the result of previous iteration. Thus, the majorized problem of \eqref{prob:X_feasible} is
\begin{equation}\label{prob:majorized_feasible_xi}
  \begin{aligned}
     & \min_{\mathbf{X}} \mathcal{R}\left\{ tr\left[\left(\mathbf{X}^{(t)} \mathbf{\Theta} - \lambda_{\theta} \mathbf{X}^{(t)}\right)^{\text{H}} \mathbf{X} \right] \right\} \\
     & \;\text{s.t. } \mathbf{X} \in \mathcal{S}_{\text{X}}.
  \end{aligned}
\end{equation}
It is clear that solving problem \eqref{prob:majorized_feasible_xi} is equivalent to solving the following problem:
\begin{equation}\label{prob:majorized_feasible_xi_simplified}
  \begin{aligned}
     & \min_{\mathbf{X}} \|\mathbf{X} - \mathbf{Z}^{(t)}\|_{\text{F}}^{2} \\
     & \;\text{s.t. } \mathbf{X} \in \mathcal{S}_{\text{X}},
  \end{aligned}
\end{equation}
where $\mathbf{Z}^{(t)} = \lambda_{\theta} \mathbf{X}^{(t)} - \mathbf{X}^{(t)} \mathbf{\Theta}$, whose closed form solution is $\mathbf{X}^{(t+1)} = e^{j \arg(\mathbf{Z}^{(t)})}$.
\subsubsection{Minimizing the Sidelobe Level Regarding $\mathbf{F}$}
Suppose that the obtained solution of problem \eqref{prob:X_feasible} is not feasible. We further minimize the sidelobe level with respect to $\mathbf{F}$ in this part. By denoting $\mathbf{\Phi} = \sum_{k \in \mathbf{\Omega}_{K}} \mathbf{J}_{k}^{\text{H}} \mathbf{X}^{\text{H}} \mathbf{X} \mathbf{J}_{k}$ and its singular value decomposition (SVD) as $\mathbf{\Phi} = \mathbf{U}_{\Phi} \mathbf{\Sigma}_{\Phi} \mathbf{U}_{\Phi}^{\text{H}}$ (assuming that the singular values are arranged in descending order), the ISL becomes
\begin{equation}\label{eq:trace_wt_F}
  \mathrm{tr}(\mathbf{Z} \mathbf{\Sigma}_{\Phi} \mathbf{Z}^{\text{H}}) = \sum_{i=1}^{N} \lambda_i(\mathbf{\Phi}) \|\mathbf{z}_i\|_2^2,
\end{equation}
where $\mathbf{Z} = \mathbf{F} \mathbf{U}_{\Phi}$, $\mathbf{z}_i$ is the $i$-th row vector of $\mathbf{Z}$, and $\lambda_i(\mathbf{\Phi})$ is the $i$-th singular value of $\mathbf{\Phi}$. Since $\mathbf{\Phi}$ is positive semidefinite, we obtain the minimum of \eqref{eq:trace_wt_F} by setting all rows of $\mathbf{Z}$ as zero except for the row corresponding to the minimum singular value of $\mathbf{\Phi}$. That is to say, we obtain an $\mathbf{F}$ that minimizes the ISL by setting $\mathbf{F} = \mathbf{\Sigma}_{\text{F}} \mathbf{U}_{\Phi}^{\text{H}}$, where $\mathbf{\Sigma}_{\text{F}} = \mathrm{Diag}(\mathbf{d}_{\text{F}})$ with $\mathbf{d}_{\text{F}} = [\mathbf{0}_{N_{\text{T}} - 1}^{\text{T}}, \sqrt{P_{\text{F}}}]^{\text{T}}$, and $\mathrm{Diag}(\mathbf{d}_{\text{F}})$ is a diagonal matrix with its main diagonal entries being $\mathbf{d}_{\text{F}}$.

\subsection{BSUM Algorithm for Inexactly Solving Subproblem \eqref{prob:sub_problem_ALM}}\label{sec:BSUM_details}
In this subsection, we propose a BSUM scheme to solve the ALM subproblem \eqref{prob:sub_problem_ALM} inexactly. In particular, the initial point at the $\ell$-th iteration is chosen as follows:
\begin{equation}\label{eq:init_point_at_lth_ALg1}
  \bm{z}_{\text{init}}^{(\ell)} =  \begin{cases}
    \bm{z}^{(\text{feas})}, & \text{if } f(\mathbf{F}^{(\text{feas})}, \mathbf{X}^{(\text{feas})}) < \mathcal{L}_{\rho^{(\ell)}}(\bm{z}^{(\ell)}; \bm{U}^{(\ell)}); \\
    \bm{z}^{(\ell)},        & \text{otherwise},
  \end{cases}
\end{equation}
where $\bm{z}^{(\ell)}$ is the approximate solution obtained at the $(\ell-1)$-th iteration of Algorithm \ref{alg:RALM}, and $\bm{z}^{(\text{feas})}$ is a feasible point obtained from Section \ref{sec:feasible_pt_details}. It follows from \eqref{eq:ALF_origin} and the definition of $\zeta$ in \eqref{eq:zeta_the_upper_bound} that
\begin{equation}
  \mathcal{L}_{\rho^{(\ell)}}(\bm{z}^{(\text{feas})}; \bm{U}^{(\ell)}) \leq f(\mathbf{F}^{(\text{feas})}, \mathbf{X}^{(\text{feas})}) \leq \zeta.
\end{equation}
This shows that the choice of the initial point in \eqref{eq:init_point_at_lth_ALg1} guarantees $\mathcal{L}_{\rho^{(\ell)}}(\bm{z}_{\text{init}}^{(\ell)}; \bm{U}^{(\ell)}) \leq \zeta$. Moreover, the proposed BSUM method has a sufficient descent property, which will be proved in Section \ref{sec:convergence_analysis}. Therefore, for any $\ell \geq 0$, we always have
\begin{equation}
  \mathcal{L}_{\rho^{(\ell)}}(\bm{z}^{(\ell+1)}; \bm{U}^{(\ell)}) \\
  \leq \mathcal{L}_{\rho^{(\ell)}}(\bm{z}_{\text{init}}^{(\ell)}; \bm{U}^{(\ell)}) \leq \zeta.
\end{equation}
Consequently, the uniform upper bound requirement in \eqref{ineq:upper_bound_requirement} can always be satisfied under the initial point choice strategy of \eqref{eq:init_point_at_lth_ALg1}.

Denote $\bm{z}^{(t)} = (\bm{C}^{(t)}, \mathbf{F}^{(t)}, \mathbf{X}^{(t)})$. Next, we present the update of each block of variables using the BSUM method.
\subsubsection{Solving for Auxiliary Variables}
The update of auxiliary variables in $\bm{C}$ can be formulated as $|\mathbf{\Omega}_{K}|$ independent optimization problems for each $\mathbf{C}_{k}$. Instead of solving \eqref{prob:sub_problem_ALM} with respect to (w.r.t.) each $\mathbf{C}_{k}$ directly, we consider the following subproblem:
\begin{align}\label{prob:block_aux_c}
   & \min_{\mathbf{C}_{k}} \frac{\rho^{(\ell)}}{2} \left\| \mathbf{X}^{(t)} \mathbf{J}_{k} {\mathbf{F}^{(t)}}^{\text{H}} - \mathbf{C}_{k} + \frac{\mathbf{U}_{k}^{(\ell)}}{\rho^{(\ell)}}\right\|_{\text{F}}^2 + \frac{\beta}{2} \left\|\mathbf{C}_{k} - \mathbf{C}_{k}^{(t)}\right\|_{\text{F}}^2, \nonumber \\
   & \text{ s.t. } \|\mathbf{C}_{k}\|_{\text{F}} \leq \xi,
\end{align}
where $\beta > 0$. As mentioned in \cite{Hallak2023}, by setting $\beta > 0$, the distance between two consecutive iterations of each $\mathbf{C}_{k}$ becomes controllable, and every update of $\mathbf{C}_{k}$ achieves a sufficient decrease. It is clear that problem \eqref{prob:block_aux_c} has a closed-form solution as follows:
\begin{equation}\label{eq:update_C}
  \mathbf{C}_{k}^{(t+1)} = \begin{cases}
    \tilde{\mathbf{C}}_{k},                                                   & \text{if } \|\tilde{\mathbf{C}}_{k}\|_{\text{F}} \leq \xi; \\
    \frac{\xi}{\|\tilde{\mathbf{C}}_{k}\|_{\text{F}}} \tilde{\mathbf{C}}_{k}, & \text{otherwise},
  \end{cases}
\end{equation}
where $\tilde{\mathbf{C}}_{k} = \frac{1}{\rho^{(\ell)} + \beta} \left(\rho^{(\ell)}\mathbf{X}^{(t)} \mathbf{J}_{k} {\mathbf{F}^{(t)}}^{\text{H}} + \mathbf{U}_{k}^{(\ell)} + \beta \mathbf{C}_{k}^{(t)}\right)$.

\subsubsection{Solving for Radar Receive Filter}\label{sec:sol_F}
The problem of solving \eqref{prob:sub_problem_ALM} w.r.t. $\mathbf{F}$ can be rewritten as
\begin{equation}\label{prob:PSL_F_1}
  \begin{aligned}
     & \min_{\mathbf{F}} \mathcal{L}_{\text{F}}(\mathbf{F})         \\
     & \;\text{s.t. } \| \mathbf{F} \|_{\text{F}}^2 = P_{\text{F}},
  \end{aligned}
\end{equation}
where $\mathcal{L}_{\text{F}}(\mathbf{F})$ is defined as follows:
\begin{multline}\label{eq:LAF_F_PSL}
  \mathcal{L}_{\text{F}}(\mathbf{F}) = \mathrm{tr}(\mathbf{F} \mathbf{Q}  \mathbf{F}^{\text{H}}) - 2 \alpha \mathcal{R}\{\mathrm{tr}(\mathbf{R}_{\text{d}}^{\text{H}} \mathbf{X} \mathbf{F}^{\text{H}})\} \\
  - \sum_{k \in \mathbf{\Omega}_{K}} \mathcal{R}\{tr[(\rho \mathbf{C}_{k} -  \mathbf{U}_{k})^{\text{H}} \mathbf{X} \mathbf{J}_{k} \mathbf{F}^{\text{H}}]\},
\end{multline}
which is reformulated from \eqref{eq:ALF_origin} by ignoring the terms irrelevant to $\mathbf{F}$ and denoting $\mathbf{Q} = \alpha \mathbf{X}^{\text{H}} \mathbf{X} + \sum_{k \in \mathbf{\Omega}_{K}} \frac{\rho}{2}\mathbf{J}_{k}^{\text{H}} \mathbf{X}^{\text{H}} \mathbf{X} \mathbf{J}_{k}$, and the superscripts of $\mathbf{X}^{(t)}$, $\bm{C}^{(t+1)}$, $\bm{U}^{(\ell)}$, and $\rho^{(\ell)}$ are omitted for notational simplicity in this part. The quadratic term $\mathrm{tr}(\mathbf{F} \mathbf{Q}  \mathbf{F}^{\text{H}})$ can be majorized at $\mathbf{F}^{(t)}$ through \textit{Lemma \ref{lm:MM}} as follows:
\begin{equation}\label{eq:F_2_MM}
  \mathrm{tr}(\mathbf{F} \mathbf{Q} \mathbf{F}^{\text{H}}) \leq \lambda_{q} \|\mathbf{F}\|_{\text{F}}^2  + 2 \mathcal{R} \{tr[(\mathbf{F}^{(t)}(\mathbf{Q} - \lambda_{q}\mathbf{I}))^{\text{H}}\mathbf{F}]\} + C_{\text{F}},
\end{equation}
where $\lambda_{q} > \|\mathbf{Q}\|_{1}$, and $C_{\text{F}}$ is the term irrelevant to $\mathbf{F}$.

By substituting \eqref{eq:F_2_MM} into \eqref{eq:LAF_F_PSL}, we have $\mathcal{G}_{\text{F}}(\mathbf{F}\mid\mathbf{F}^{(t)})$ as the tight upper bound of $\mathcal{L}_{\text{F}}(\mathbf{F})$, i.e., $\mathcal{G}_{\text{F}}(\mathbf{F}\mid\mathbf{F}^{(t)}) \geq \mathcal{L}_{\text{F}}(\mathbf{F})$ for all $\mathbf{F} \in \mathcal{S}_{\text{F}}$ with equality holding when $\mathbf{F} = \mathbf{F}^{(t)}$, and
\begin{equation}\label{eq:neg_F_MM2}
  \mathcal{G}_{\text{F}}(\mathbf{F}\mid\mathbf{F}^{(t)}) = \lambda_{q} \|\mathbf{F}\|_{\text{F}}^2 - 2 \mathcal{R} \left\{tr \left( {\mathbf{\Xi}^{(t)}}^{\text{H}} \mathbf{F} \right)\right\} + C_{\text{F}},
\end{equation}
where
\begin{equation}
  \mathbf{\Xi}^{(t)} = \sum_{k \in \mathbf{\Omega}_{K}} \frac{1}{2}(\rho \mathbf{C}_{k} - \mathbf{U}_{k})^{\text{H}} \mathbf{X} \mathbf{J}_{k} + \alpha \mathbf{R}_{\text{d}}^{\text{H}} \mathbf{X} - \mathbf{F}^{(t)} \mathbf{Q} + \lambda_{q} \mathbf{F}^{(t)}.
\end{equation}
Then, the majorized problem of \eqref{prob:PSL_F_1} is
\begin{equation}\label{prob:PSL_F_MMed_lg0}
  \begin{aligned}
     & \min_{\mathbf{F}} \mathcal{G}_{\text{F}}(\mathbf{F}\mid\mathbf{F}^{(t)}) \\
     & \;\text{s.t. } \| \mathbf{F} \|_{\text{F}}^2 = P_{\text{F}}.
  \end{aligned}
\end{equation}
Problem \eqref{prob:PSL_F_MMed_lg0} has a closed-form solution as follows:
\begin{equation}\label{sol:F}
  \mathbf{F}^{(t+1)} = \frac{\sqrt{P_{\text{F}}}}{\| \mathbf{\Xi}^{(t)} \|_{\text{F}}}\mathbf{\Xi}^{(t)}.
\end{equation}
\subsubsection{Solving for Transmit Waveform}
The problem of solving \eqref{prob:sub_problem_ALM} w.r.t. $\mathbf{X}$ can be rewritten as
\begin{equation}\label{prob:X}
  \begin{aligned}
     & \min_{\mathbf{X}} \mathcal{L}_{\text{X}} (\mathbf{X}), \\
     & \; \text{s.t. } \mathbf{X} \in \mathcal{S}_{\text{X}},
  \end{aligned}
\end{equation}
where
\begin{multline}\label{eq:ALF_wrt_X}
  \mathcal{L}_{\text{X}}(\mathbf{X}) = \mathrm{tr}(\mathbf{X} \mathbf{P} \mathbf{X}^{\text{H}}) - 2 \alpha \mathcal{R}\{\mathrm{tr}(\mathbf{X} \mathbf{F}^{\text{H}} \mathbf{R}_{\text{d}}^{\text{H}})\} \\
  - \sum_{k \in \mathbf{\Omega}_{K}} \mathcal{R}\{tr[(\rho \mathbf{C}_{k} -  \mathbf{U}_{k})^{\text{H}} \mathbf{X} \mathbf{J}_{k} \mathbf{F}^{\text{H}}]\}\\
  + (1 - \alpha) (\mathbf{x}^{\text{H}} \mathbf{H}^{\text{H}} \mathbf{H} \mathbf{x} - 2 \mathcal{R}\{\mathbf{x}^{\text{H}} \mathbf{H}^{\text{H}} \mathbf{s}\}),
\end{multline}
and $\mathbf{P} = \alpha {\mathbf{F}^{(t+1)}}^{\text{H}} \mathbf{F}^{(t+1)} + \frac{\rho}{2} \sum_{k \in \mathbf{\Omega}_{K}} \mathbf{J}_{k} {\mathbf{F}^{(t+1)}}^{\text{H}} \mathbf{F}^{(t+1)} \mathbf{J}_{k}^{\text{H}}$.

We consider majorizing the two quadratic terms, $\mathrm{tr}(\mathbf{X} \mathbf{P} \mathbf{X}^{\text{H}})$ and $\mathbf{x}^{\text{H}} \mathbf{H}^{\text{H}} \mathbf{H} \mathbf{x}$ separately. According to \textit{Lemma \ref{lm:MM}}, we have
\begin{equation}\label{eq:mm_1_X}
  \mathrm{tr}(\mathbf{X} \mathbf{P} \mathbf{X}^{\text{H}}) \leq \lambda_{p} \|\mathbf{X}\|_{\text{F}}^2 + 2 \mathcal{R}\{\mathrm{tr}(\mathbf{X}^{(t)} \mathbf{P} - \lambda_{p} \mathbf{X}^{(t)})^{\text{H}} \mathbf{X}\} + C_{\text{X}},
\end{equation}
where $\lambda_{p} > \|\mathbf{P}\|_{1}$, and $C_{\text{X}}$ is the term irrelevant to $\mathbf{X}$. Similarly,
\begin{equation}\label{eq:mm_2_x}
  \mathbf{x}^{\text{H}} \mathbf{H}^{\text{H}} \mathbf{H} \mathbf{x} \leq \lambda_{h} \|\mathbf{x}\|_{2}^{2} + 2 \mathcal{R} \{\mathbf{x}^{\text{H}} (\mathbf{H}^{\text{H}} \mathbf{H} - \lambda_h \mathbf{I}) \mathbf{x}^{(t)}\} + C_{\text{x}},
\end{equation}
in which $\lambda_{h} = \lambda_{\max}(\mathbf{H}^{\text{H}} \mathbf{H})$ and $C_{\text{x}}$ is irrelevant to $\mathbf{x}$. By substituting \eqref{eq:mm_1_X} and \eqref{eq:mm_2_x} into \eqref{eq:ALF_wrt_X}, the tight upper bound of $\mathcal{L}_{\text{X}}(\mathbf{X})$ is
\begin{equation}
  \mathcal{G}_{\text{X}}(\mathbf{X} | \mathbf{X}^{(t)}) = [\lambda_{p} + (1 - \alpha) \lambda_{h}] \|\mathbf{X}\|_{\text{F}}^2 - 2 \mathcal{R}\{\mathrm{tr}[{\mathbf{\Psi}^{(t)}}^{\text{H}} \mathbf{X}]\} + \tilde{C}_{\text{x}},
\end{equation}
where $\mathbf{\Psi}^{(t)} = \alpha \mathbf{R}_{\text{d}} \mathbf{F} + \sum_{k \in \mathbf{\Omega}_{K}} \frac{1}{2} (\rho \mathbf{C}_{k} - \mathbf{U}_{k}) \mathbf{F} \mathbf{J}_{k}^{\text{H}} - (1 - \alpha) {\mathbf{\Phi}^{(t)}}^{\text{T}} - \mathbf{X}^{(t)} (\mathbf{P} - \lambda_{p} \mathbf{I})$ with $\mathbf{\Phi}^{(t)} = \mathrm{mat}\{(\mathbf{H}^{\text{H}} \mathbf{H} - \lambda_h \mathbf{I}) \mathbf{x}^{(t)} - \mathbf{H}^{\text{H}} \mathbf{s}\}$, $\mathrm{mat}\{\cdot\}$ means reshaping a column vector into a matrix, and $\tilde{C}_{\text{x}} = (1 - \alpha) C_{\text{x}} + C_{\text{X}}$. Then, the majorized problem of \eqref{prob:X} is
\begin{equation}
  \begin{aligned}
     & \min_{\mathbf{X}} \mathcal{G}_{\text{X}}(\mathbf{X} | \mathbf{X}^{(t)}) \\
     & \; \text{s.t. } \mathbf{X} \in \mathcal{S}_{\text{X}},
  \end{aligned}
\end{equation}
whose closed-form solution is
\begin{equation}\label{sol:X}
  \mathbf{X}^{(t+1)} = e^{j \mathrm{arg}(\mathbf{\Psi}^{(t)})},
\end{equation}

The proposed BSUM algorithm for solving the ALM problem \eqref{prob:sub_problem_ALM} inexactly is summarized in Algorithm \ref{alg:inexact_ALM_sol}. It is worth mentioning that the SQUAREM scheme \cite{varadhan2008simple} is implemented when updating $\mathbf{F}$ and each row of $\mathbf{X}$ in Algorithm \ref{alg:inexact_ALM_sol} to accelerate the convergence. The BSUM algorithm performs efficiently since every variable admits closed-form updates. The computational cost of updating $\mathbf{F}$, $\mathbf{X}$, $\bm{C}$, and $\bm{U}$ is dominated by calculating $\mathbf{\Xi}$, $\mathbf{\Psi}$, and $\mathbf{X} \mathbf{J}_{k} \mathbf{F}^{\text{H}}$ for each $k \in \mathbf{\Omega}_{K}$, respectively, whose computational complexities are $\mathcal{O}(|\mathbf{\Omega}_{K}| \cdot [(N_{\text{T}} + T) N_{\text{R}} T])$, $\mathcal{O}(|\mathbf{\Omega}_{K}| \cdot [(N_{\text{R}} + T) N_{\text{T}} T])$, $\mathcal{O}(|\mathbf{\Omega}_{K}| \cdot [(N_{\text{R}} + T) N_{\text{T}} T])$, and $\mathcal{O}(|\mathbf{\Omega}_{K}| \cdot [(N_{\text{R}} + T) N_{\text{T}} T])$, respectively. Therefore, the total per-iteration complexity of Algorithm \ref{alg:inexact_ALM_sol} is $\mathcal{O}(|\mathbf{\Omega}_{K}| \cdot [(N_{\text{T}}T +  N_{\text{R}} T +  N_{\text{T}}  N_{\text{R}} ) T])$, which scales linearly with the number of range bins for sidelobe suppression $|\mathbf{\Omega}_{K}|$, the number of transmit antennas $N_{\text{T}}$, and the number of receive antennas $N_{\text{R}}$.
\begin{algorithm}[t]
  \caption{BSUM for Inexactly Solving Problem \eqref{prob:sub_problem_ALM}}
  \begin{algorithmic}[1]\label{alg:inexact_ALM_sol}
    \REQUIRE Error tolerance $\varepsilon^{(\ell)}$ and $\rho^{(\ell)}$ from Algorithm \ref{alg:RALM}.
    \REPEAT
    \FOR{$k \in \mathbf{\Omega}_{K}$}
    \STATE Update $\mathbf{C}_{k}^{(t+1)}$ by \eqref{eq:update_C};
    \ENDFOR
    \STATE Update $\mathbf{F}^{(t+1)}$ by \eqref{sol:F};
    \STATE Update $\mathbf{X}^{(t+1)}$ by \eqref{sol:X}
    \STATE $t \gets t+1$;
    \UNTIL{$(\bm{C}^{(t)}, \mathbf{F}^{(t)}, \mathbf{X}^{(t)})$ is an $\varepsilon^{(\ell)}$-stationary point of problem \eqref{prob:sub_problem_ALM}};
    \ENSURE $(\bm{C}^{(t)}, \mathbf{F}^{(t)}, \mathbf{X}^{(t)})$.
  \end{algorithmic}
\end{algorithm}
\subsection{Convergence Analysis}\label{sec:convergence_analysis}
The convergence analysis of the proposed inexact ALM algorithm (i.e., Algorithm \ref{alg:RALM}) consists of four parts. We first show the sequence of the variables generated by Algorithm \ref{alg:inexact_ALM_sol} enjoys a sufficient descent property in Lemma \ref{lm:2}; then, we show that the subgradient of $\mathcal{L}_{\rho}(\bm{C}, \mathbf{F}, \mathbf{X}; \bm{U})$ is bounded in Lemma \ref{lm:3}; based on the results of Lemma \ref{lm:2} and Lemma \ref{lm:3}, we show that Algorithm \ref{alg:inexact_ALM_sol} achieves an $\varepsilon$-stationary point within a finite number of iterations in Theorem \ref{thm:1}; finally, we prove that any limit point generated by Algorithm \ref{alg:RALM} is a feasible stationary point in Theorem \ref{thm:2}.

The convergence analysis relies significantly on the uniform boundness of the primal variables $(\bm{C}, \mathbf{F}, \mathbf{X})$ and the multiplier $\bm{U}$ generated by Algorithm \ref{alg:inexact_ALM_sol}. The boundness of $\bm{U}$ can be guaranteed by the updating rule proposed in \eqref{eq:uk_update_bound}, and the boundness of $(\bm{C}, \mathbf{F}, \mathbf{X})$ can be ensured by the updating rules proposed in \eqref{eq:update_C}, \eqref{sol:F}, and \eqref{sol:X}.

\begin{lemmaNoParens}[(Sufficient descent property)]\label{lm:2}
  At the $\ell$-th iteration of Algorithm \ref{alg:RALM}, there exist positive $\tau_{f}$ and $\tau_{x}$ such that
  \begin{multline}
    \mathcal{L}_{\rho^{(\ell)}}(\bm{z}^{(t)}; \bm{U}^{(\ell)}) - \mathcal{L}_{\rho^{(\ell)}}(\bm{z}^{(t+1)}; \bm{U}^{(\ell)}) \geq \tau_{x} \|\mathbf{X}^{(t)} - \mathbf{X}^{(t+1)}\|_{\text{F}}^2                                                                                                                                                          \\
    + \tau_{f} \|\mathbf{F}^{(t)} - \mathbf{F}^{(t+1)}\|_{\text{F}}^2 + \frac{\beta}{2} \sum_{k \in \mathbf{\Omega}_{K}} \|\mathbf{C}_{k}^{(t)} - \mathbf{C}_{k}^{(t+1)}\|_{\text{F}}^2.
  \end{multline}
\end{lemmaNoParens}
\begin{proof}
  See Appendix \ref{proof:lemma2}.
\end{proof}

\begin{lemmaNoParens}[(Subgradient boundness)]\label{lm:3}
  At the $\ell$-th iteration of Algorithm \ref{alg:RALM}, there exist a subgradient $\bm{J}^{(t+1)} = (\mathbf{J}_{C}^{(t+1)}, \mathbf{J}_{\rm F}^{(t+1)}, \mathbf{J}_{\rm X}^{(t+1)})$ with $\bm{J}^{(t+1)} \in \partial [\mathcal{L}_{\rho^{(\ell)}}(\bm{z}^{(t+1)}; \bm{U}^{(\ell)}) + \mathbb{I}_{\mathcal{S}}(\bm{z}^{(t+1)})]$ and a constant $M>0$ such that
  \begin{equation}\label{ineq:lm3_subgrad_upperbd}
    \begin{aligned}
      \|\bm{J}^{(t+1)}\|_{{\rm F}} \leq M (\|\mathbf{F}^{(t)} & - \mathbf{F}^{(t+1)}\|_{{\rm F}}  + \|\mathbf{X}^{(t)} - \mathbf{X}^{(t+1)}\|_{{\rm F}}          \\
                                                              & + \sum_{k \in \mathbf{\Omega}_{K}} \|\mathbf{C}_{k}^{(t)} - \mathbf{C}_{k}^{(t+1)}\|_{{\rm F}}).
    \end{aligned}
  \end{equation}
\end{lemmaNoParens}
\begin{proof}
  See Appendix \ref{proof:lemma3}.
\end{proof}

\begin{theorem}[(Iteration complexity to obtain an $\varepsilon^{(\ell)}$-stationary point)]\label{thm:1}
  Given any positive $\epsilon^{(\ell)}$ and $\rho^{(\ell)}$, Algorithm \ref{alg:inexact_ALM_sol} returns an $\varepsilon^{(\ell)}$-stationary point in $\mathcal{O} \left(\frac{{\rho^{(\ell)}}^2}{{\varepsilon^{(\ell)}}^2}\right)$ iterations.
\end{theorem}
\begin{proof}
  See Appendix \ref{proof:thm1}.
\end{proof}

\begin{theorem}[(Convergence to the stationary point)]\label{thm:2}
  Any limit point of $\{(\mathbf{F}^{(\ell+1)}, \mathbf{X}^{(\ell+1)})\}_{\ell \geq 0}$ generated by Algorithm \ref{alg:RALM} is a stationary point of problem \eqref{prob:prob_origin}.
\end{theorem}
\begin{proof}
  See Appendix \ref{proof:thm2}.
\end{proof}
\section{Numerical Results}\label{sec:num_res}
In this section, we present simulation results to evaluate the performance of the proposed algorithms under various parameter configurations. The communication channel $\mathbf{H}$ is generated by following \cite{932706} with Extended Pedestrian A (EPA) fading profile \cite{3gpp.36.104}. The information symbol matrix $\mathbf{S}$ is modulated using a unit-power QPSK alphabet, with each entry in $\mathbf{S}$ having unit power. The desired radar receive spatial response, $\mathbf{R}_{\text{d}}$, is considered to be a $3$ dB beamwidth of $20^{\circ}$ focusing at $0^{\circ}$, generated by the algorithm in \cite{li2007mimo}. It is worth noting that the algorithm proposed in this paper can be directly extended to the Extremely large-scale MIMO (XL-MIMO) system \cite{10500425} for generating unimodular low sidelobe level ISAC sequences.
\begin{table}[t]
  \centering
  \caption{Parameters setting in simulations.}\label{tab:parameters}
  \begin{tabular}{|c|c|c|}
    \hline Parameter      & Definition                                                                        & Value     \\
    \hline $N_{\text{T}}$ & Number of transmit antennas                                                       & 8         \\
    \hline $N_{\text{R}}$ & Number of radar receive antennas                                                  & 8         \\
    \hline $N_{\text{C}}$ & Number of users                                                                   & 4         \\
    \hline $L$            & CP length                                                                         & 6         \\
    \hline $T$            & Block length                                                                      & 64        \\
    \hline $K$            & \makecell[c]{Maximum index of range bin of interest}                              & 6         \\
    \hline $\xi^{\prime}$ & Desired PSLR, $\xi = \sqrt{T N_{\text{T}}} \cdot 10^{(-\frac{\xi^{\prime}}{20})}$ & 30 dB     \\
    \hline $P_{\text{F}}$ & \makecell[c]{Power of the receive filter}                                         & 64        \\
    \hline $\delta$       & Violation of constraints descent criterion in \eqref{eq:rho_update}               & 0.965     \\
    \hline $\gamma$       & Penalty parameter update factor in \eqref{eq:rho_update}                          & 1.1       \\
    \hline $\beta$        & Parameter in \eqref{prob:block_aux_c}                                             & 1         \\
    \hline $u_{\max}$     & Upper bound of Lagrange multiplier                                                & $10^{3}$  \\
    \hline $\rho^{(0)}$   & Initial penalty parameter                                                         & $10^{-3}$ \\
    \hline $-$            & Maximum inner iterations                                                          & $50$      \\
    \hline
  \end{tabular}
\end{table}

The stopping criterion for Algorithm \ref{alg:inexact_ALM_sol} at the $\ell$-th outer loop is set as $\varepsilon^{(\ell)} = \sup_{\bm{z}^{(\ell)}} \|\bm{J}^{(0)}\|_{\text{F}}/\ell$, where $\sup_{\bm{z}^{(\ell)}} \|\bm{J}^{(0)}\|_{\text{F}}$ is the subgradient upper bound in \eqref{ineq:lm3_subgrad_upperbd}. Unless otherwise specified, the parameter configurations are listed in Table \ref{tab:parameters}. Algorithm \ref{alg:RALM} terminates after 500 iterations or when the following optimality violation criterion is met
\begin{equation}\label{eq:criteria}
  \max\{e^{(\ell)}, v^{(\ell)}\} \leq \sqrt{T} \times 10^{-3},
\end{equation}
where $v^{(\ell)}$ is defined in \eqref{eq:constraints_violation}, and
\begin{equation}
  e^{(\ell)} = \sup_{\bm{z}^{(\ell)}} \|\mathbf{J}_{\text{F}}^{(\ell)}\|_{\text{F}} + \sup_{\bm{z}^{(\ell)}} \|\mathbf{J}_{\text{X}}^{(\ell)}\|_{\text{F}} + \sum_{k \in \mathbf{\Omega}_{K}} \sup_{\bm{z}^{(\ell)}} \|\mathbf{J}_{C_{k}}^{(\ell)}\|_{\text{F}},  \nonumber
\end{equation}
which are the summation of the upper bounds in \eqref{eq:subgrad_LF}, \eqref{eq:sub_grad_bound_X}, and \eqref{eq:sub_grad_bound_C}.

\begin{figure}[t]
  \centering
  \subfloat[The curve of the optimality violation.] {
    \includegraphics[width=0.48\columnwidth]{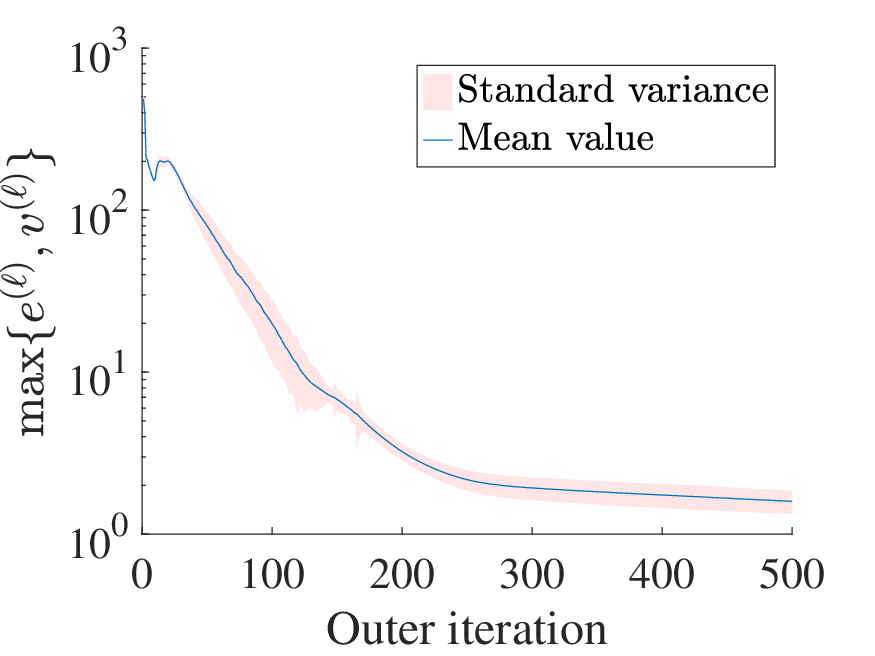}
    \label{fig:MC_subgrad_curve}
  }
  \subfloat[The curve of sum data rate and sidelobe level.] {
    \includegraphics[width=0.48\columnwidth]{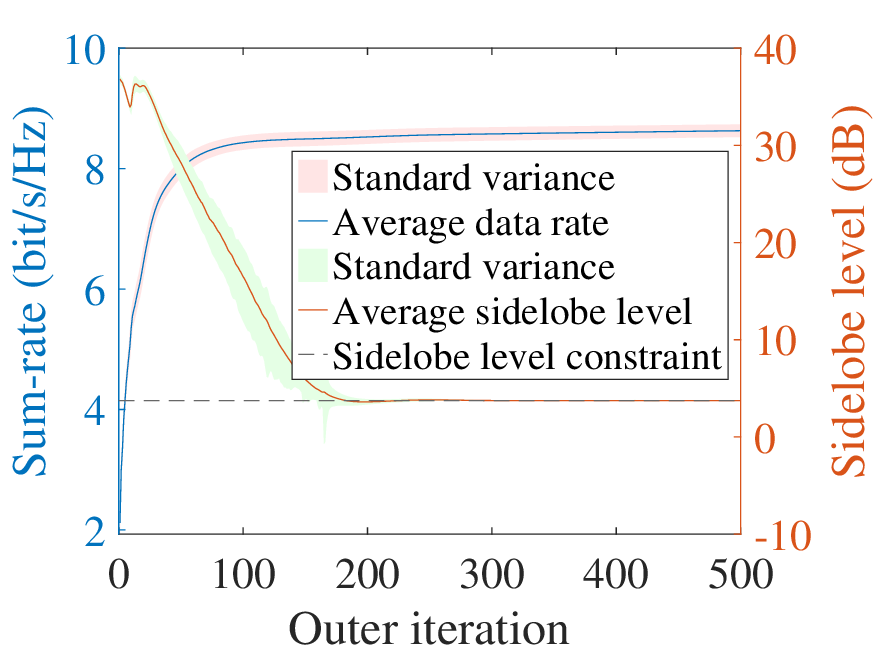}
    \label{fig:evo_rate}
  }
  \hfill
  \subfloat[The curve of BP MSE.] {
    \includegraphics[width=0.48\columnwidth]{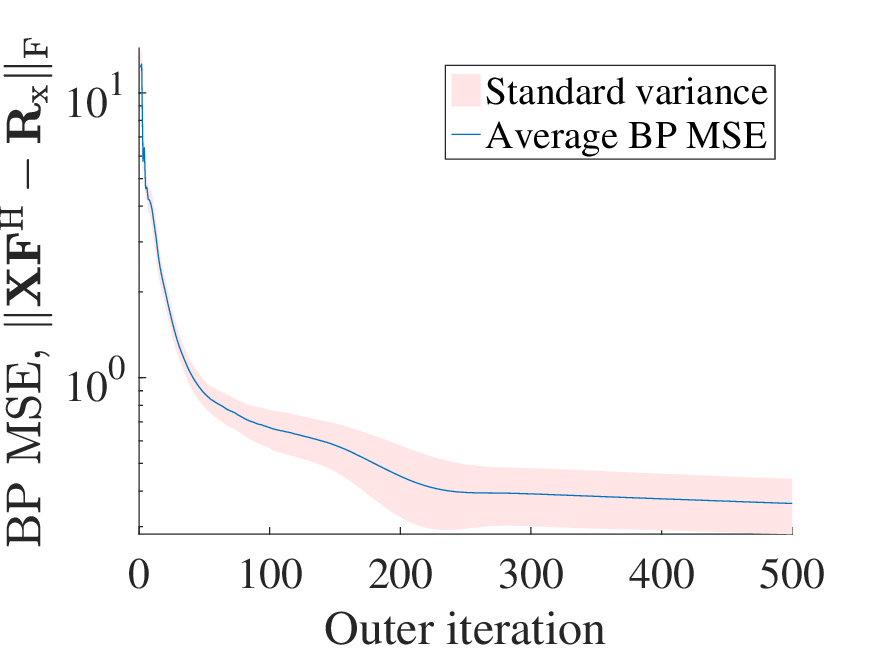}
    \label{fig:evo_bp_PSL}
  }
  \subfloat[The statistics of the number of inner iterations.] {
    \includegraphics[width=0.48\columnwidth]{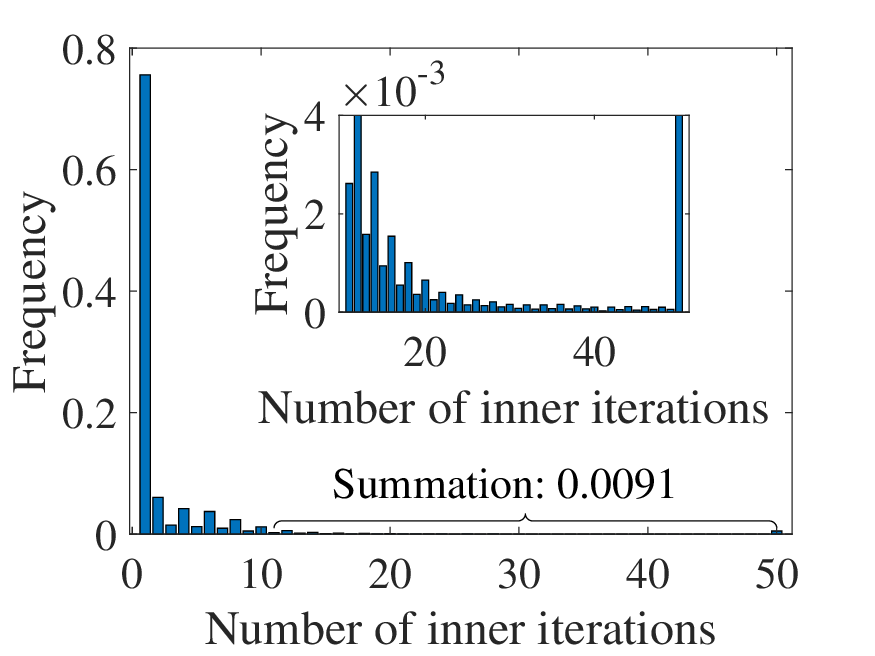}
    \label{fig:MC_inner_inter_curve}
  }
  \caption{The average iteration curves of 1000 Monte-Carlo simulations on Algorithm \ref{alg:RALM}.}
  \label{fig:MC_iterations}
\end{figure}
\subsection{Convergence Performance}
In this subsection, we evaluate the convergence behavior of the proposed inexact ALM algorithm using 1000 Monte Carlo (MC) simulations. In each simulation, the channel matrix $\mathbf{H}$ and the information symbol matrix $\mathbf{S}$ are generated independently and randomly, with the variables $(\bm{C}, \mathbf{F}, \mathbf{X}), \bm{U}$ also randomly initialized. The predetermined feasible points are consistent across all MC simulations. The shaded areas in the figures represent the standard deviation of the results from the 1000 simulations.

Fig. \ref{fig:MC_iterations} shows the average behavior of Algorithm \ref{alg:RALM}. Fig. \ref{fig:MC_iterations}\subref{fig:MC_subgrad_curve} plots the optimality violation in \eqref{eq:criteria} alongside the stopping threshold (dashed line) for Algorithm \ref{alg:RALM}. On average, the algorithm reaches the stopping criterion after approximately 480 iterations. Due to the random initialization of variables, the optimized sum rate and beampattern MSE vary. Fig. \ref{fig:MC_iterations}\subref{fig:evo_rate} and Fig. \ref{fig:MC_iterations}\subref{fig:evo_bp_PSL} illustrate the average curves of sum rate, beampattern MSE, and sidelobe level, showing that the final results perform well on average, with the maximum sidelobe level approaching the constraint. Fig. \ref{fig:MC_iterations}\subref{fig:MC_inner_inter_curve} shows the number of inner iterations required to find the $\varepsilon^{(\ell)}$-stationary point. Although the maximum allowed inner iterations is 50, over 99.5$\%$ of the inner loops converge within 10 iterations. Combined with closed-form updates and the ``nearest-vector'' algorithm, the proposed BSUM algorithm efficiently solves the ALM subproblems.

\begin{figure}[t]
  \centering
  \subfloat[The average cross-correlation level.] {
    \includegraphics[width=0.48\columnwidth]{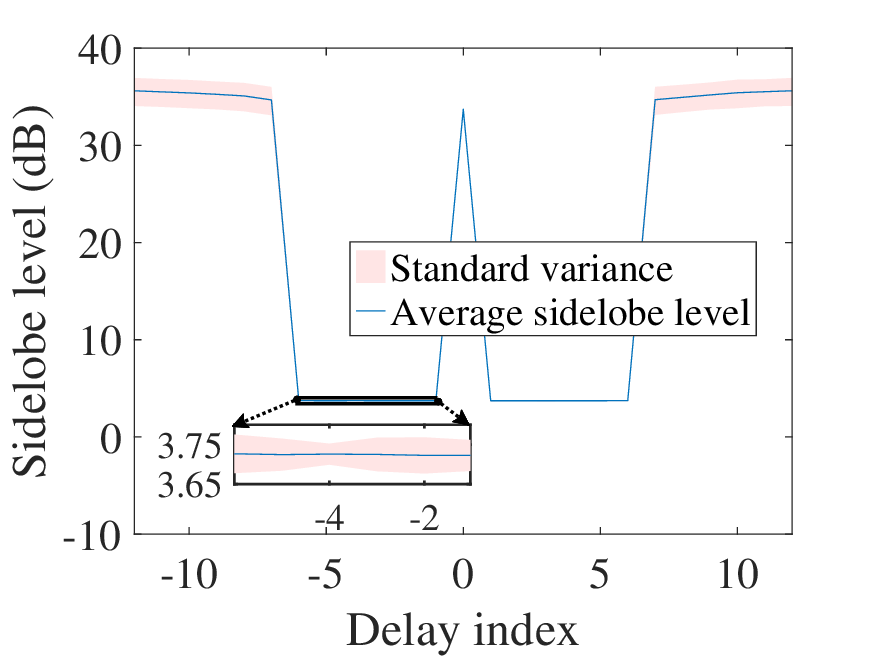}
  }
  \subfloat[The average radar receive beampattern.] {
    \includegraphics[width=0.48\columnwidth]{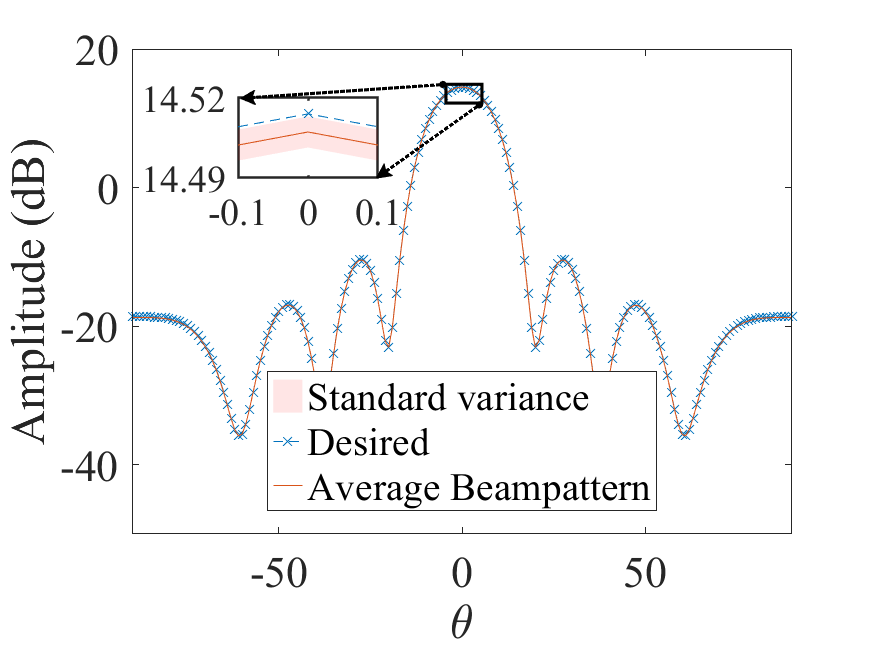}
  }
  \caption{The average results of 1000 Monte-Carlo simulations on Algorithm \ref{alg:RALM} after meeting the stopping criteria.}
  \label{fig:MC_performance}
\end{figure}

\begin{figure}[t]\centering
  \centering
  \subfloat[The cross-correlation level between $\mathbf{X}$ and $\mathbf{F}$ under different $K$.] {
    \includegraphics[width=0.48\columnwidth]{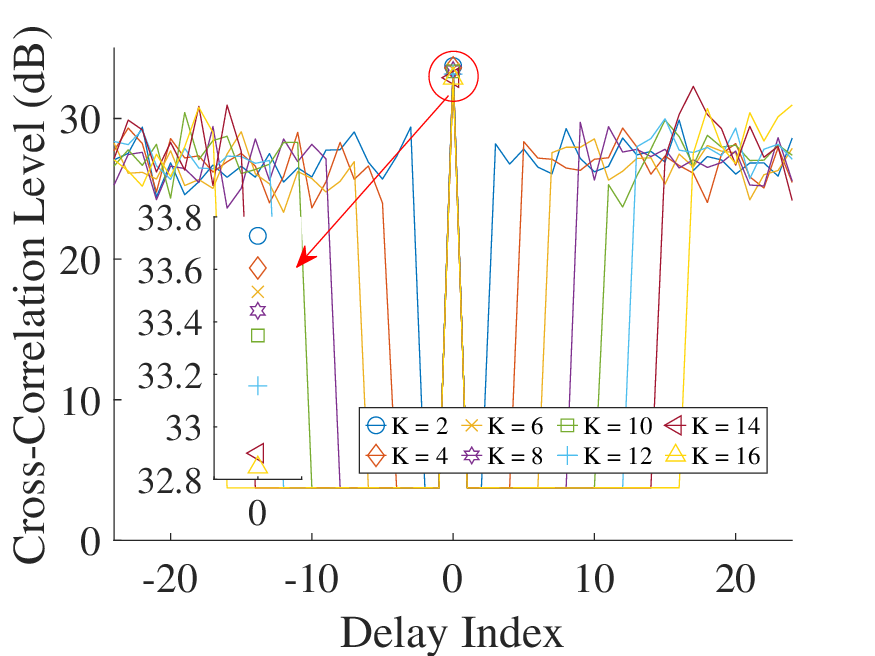}\label{fig:SL_K}
  }
  \subfloat[The sum rate under SNR = $6$ dB and beampattern MSE w.r.t $K$.] {
    \includegraphics[width=0.48\columnwidth]{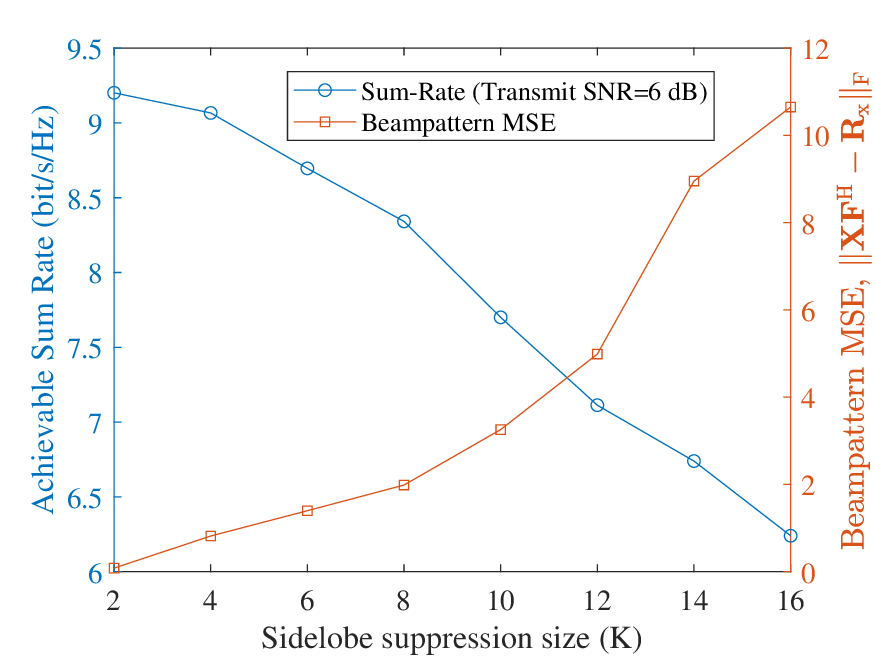}\label{fig:rate_MSE_K}
  }
  \caption{Properties of optimized ISAC waveform w.r.t. $K$ under $\alpha = 0.2$.}
  \label{fig:impact_K}
\end{figure}
Fig. \ref{fig:MC_performance} shows the average cross-correlation level and radar receive beampattern of the sequences generated by Algorithm \ref{alg:RALM} after convergence. The results indicate that all sidelobes in the interested area are well suppressed, and the optimized radar beampattern closely matches the desired one.
\subsection{Communication and Radar Performance under Different System Configurations}
In this subsection, we evaluate the communication and radar performance of the proposed algorithm under different system configurations. Specifically, we examine the impact of the range sidelobe suppression area size determined by $K$ and the maximum sidelobe level constraints $\xi^{\prime}$ (dB).

Fig. \ref{fig:impact_K} illustrates the impact of the sidelobe suppression area sizes determined by $K$ on the optimized results. As shown in Fig. \ref{fig:impact_K}\subref{fig:SL_K}, even with the increment of $K$, the optimized sidelobe levels can still approach the predetermined threshold. In Fig. \ref{fig:impact_K}\subref{fig:rate_MSE_K}, the achievable sum rate decreases, the beampattern MSE increases, and the mainlobe level decreases slightly (shown at the bottom-left of Fig. \ref{fig:impact_K}\subref{fig:SL_K}) as $K$ grows, since a larger suppression area imposes more constraints, reducing the feasible region and degrading the performance of optimized results.
\begin{figure}[t]\centering
  \centering
  \subfloat[The cross-correlation level under different PSLR constraints.] {
    \includegraphics[width=0.47\columnwidth]{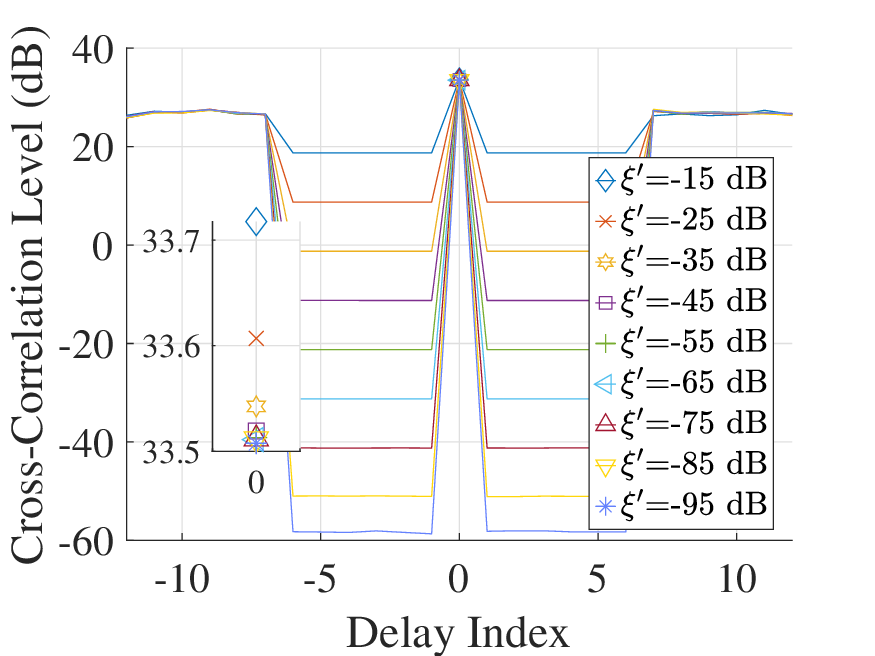}
    \label{fig:SL_PSLR}
  }
  \subfloat[The sum rate under SNR = $6$ dB and beampattern MSE w.r.t. PSLR constraints.] {
    \includegraphics[width=0.47\columnwidth]{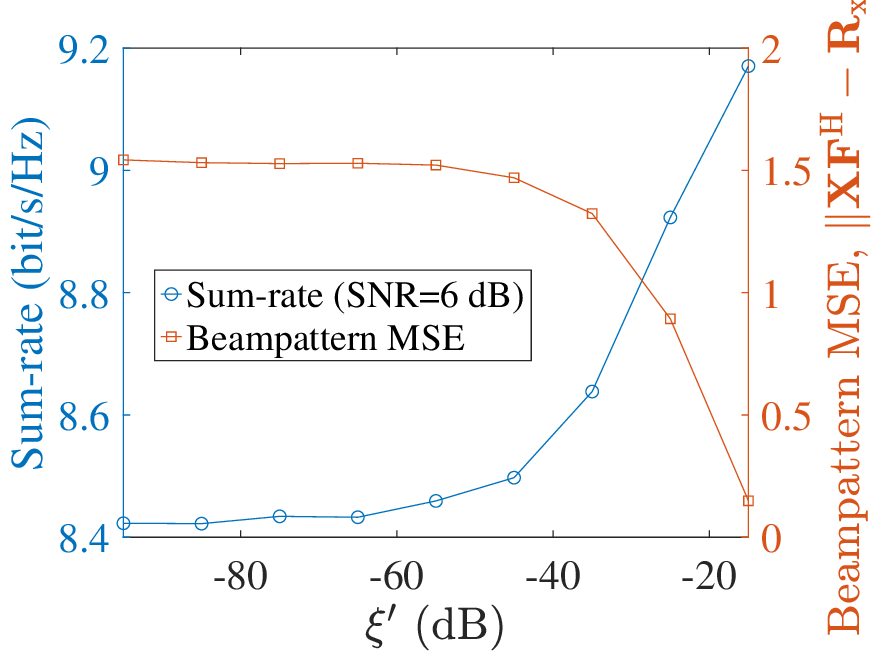}
    \label{fig:rate_MSE_PSLR}
  }
  \caption{Properties of optimized ISAC waveform with different PSLR under $\alpha = 0.2$.}
  \label{fig:impact_PSLR}
\end{figure}

Fig. \ref{fig:impact_PSLR} shows the impact of maximum sidelobe level constraints on the optimized results. As seen in Fig. \ref{fig:impact_PSLR}\subref{fig:SL_PSLR}, the sidelobe constraints can always be met. Lower sidelobe constraints require $\mathbf{X}$ and $\mathbf{F}$ to be less correlated, making the problem more complex and resulting in worse performance in both S$\&$C. This can be observed in Fig. \ref{fig:impact_PSLR}\subref{fig:rate_MSE_PSLR}, in which a lower maximum sidelobe level constraint leads to a worse achievable sum rate and a larger beampattern MSE.
\begin{figure*}[t]
  \centering
  \subfloat[The sum data rate by different algorithms.] {
    \includegraphics[width=0.65\columnwidth]{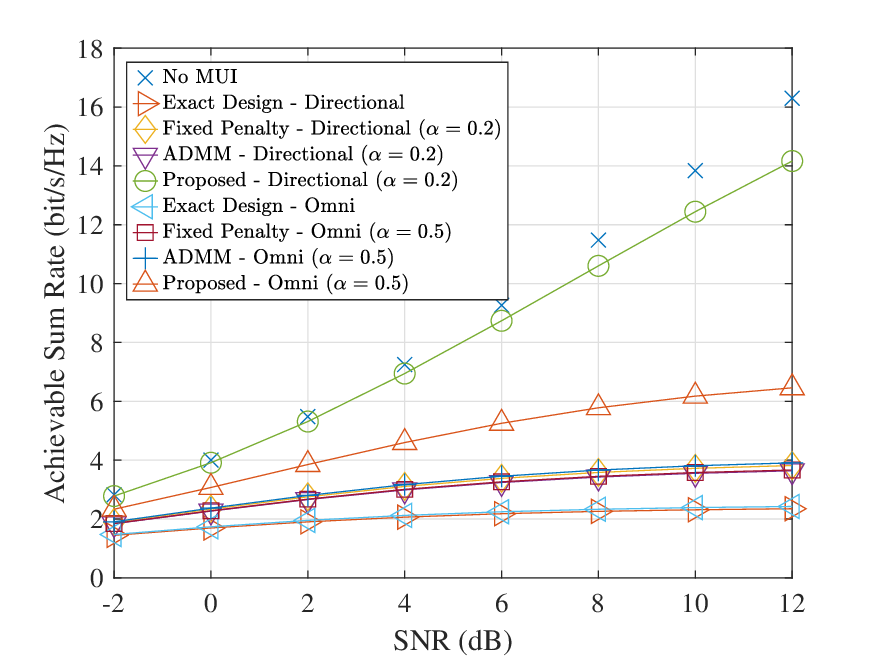}
    \label{fig:rate_algs}
  }
  \subfloat[The radar receive beampattern by different algorithms.] {
    \includegraphics[width=0.65\columnwidth]{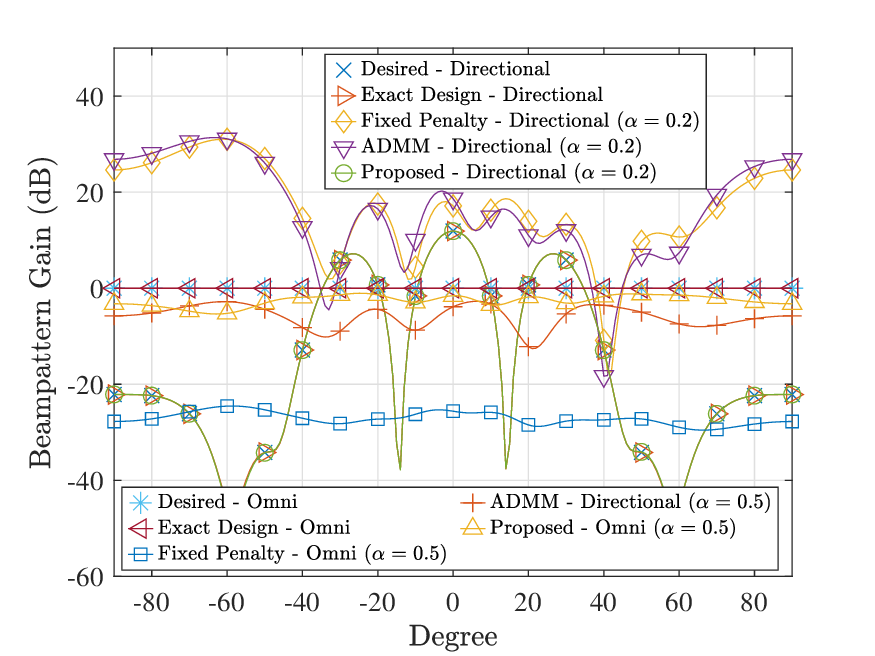}
    \label{fig:bpMSE_algs}
  }
  \subfloat[The cross-correlation level by different algorithms.] {
    \includegraphics[width=0.65\columnwidth]{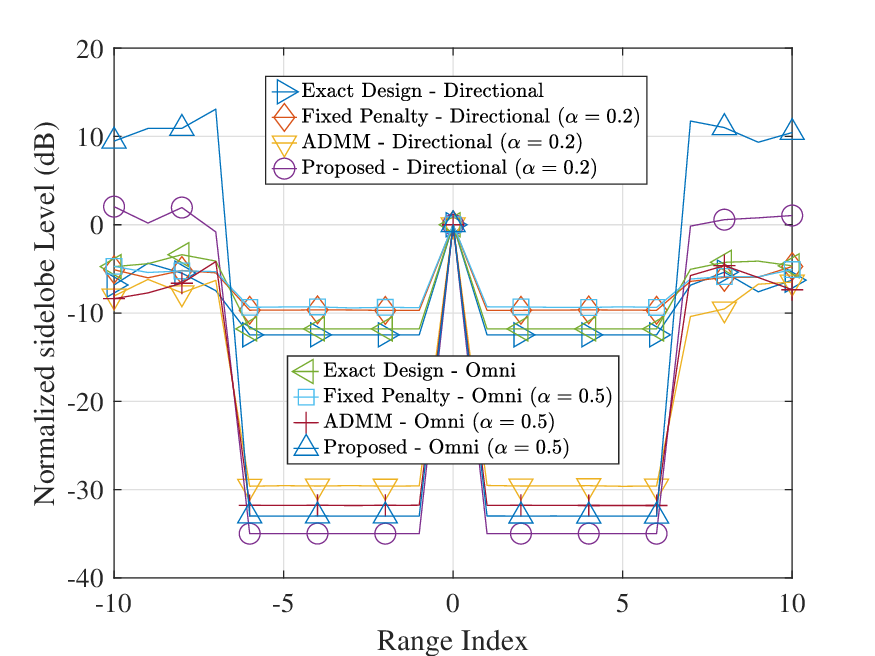}
    \label{fig:SL_algs}
  }
  \caption{The performance comparison of different algorithms.}
  \label{fig:Performance_compare}
\end{figure*}
\subsection{Comparison with Existing SOTA Approaches}
\begin{figure*}[t]
  \centering
  \subfloat[]{
    \includegraphics[width=0.18\textwidth]{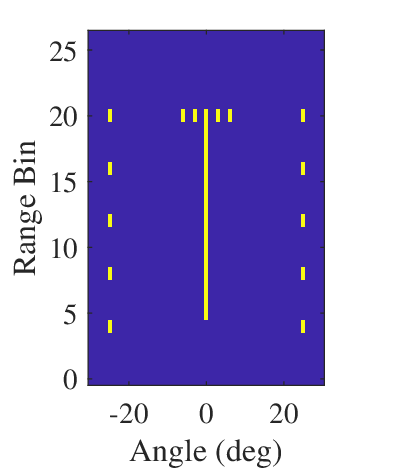}
    \label{fig:image_origin}
  }
  \subfloat[]{
    \includegraphics[width=0.18\textwidth]{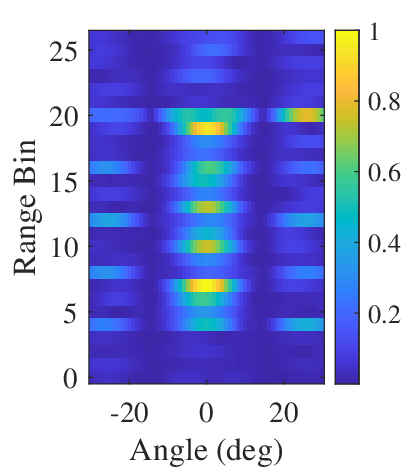}
    \label{fig:image_proposed}
  }
  % \hfill
  \subfloat[]{
    \includegraphics[width=0.18\textwidth]{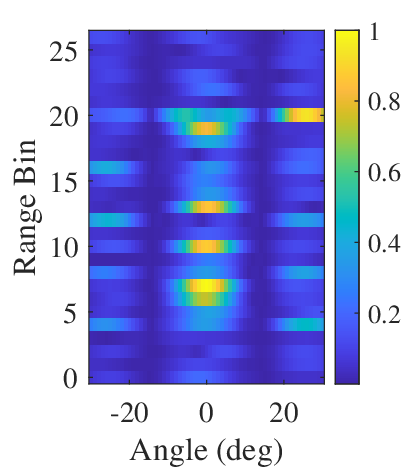}
    \label{fig:image_exact}
  }
  \subfloat[]{
    \includegraphics[width=0.18\textwidth]{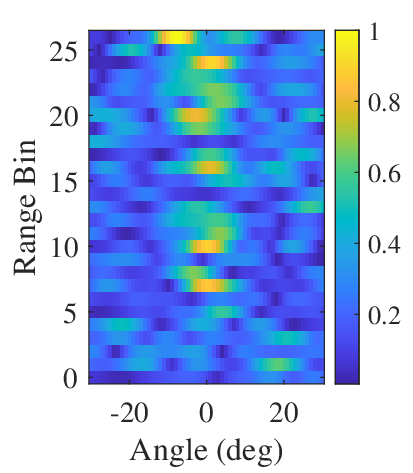}
    \label{fig:image_admm}
  }
  \subfloat[]{
    \includegraphics[width=0.18\textwidth]{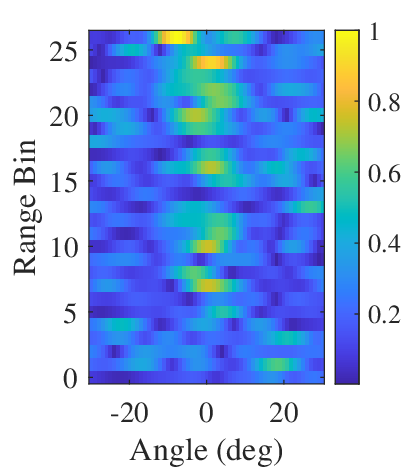}
    \label{fig:image_fixed}
  }
  \caption{The MIMO radar image under the results of different algorithms when SNR = $15$ dB: (a) Original image of ``T''. (b) The image formed by the proposed ALM algorithm. (c) The image formed by the Exact Design scheme. (d) The image formed by the fixed penalty ALM algorithm. (e) The image formed by the ADMM algorithm.}
  \label{fig:images}
\end{figure*}
In this subsection, we compare the radar and communication performance of the proposed approach with the modified work \cite{Li2008} and the ALM algorithm with a fixed penalty parameter. In \cite{Li2008}, the joint design of the radar receive filter and transmit waveform focuses on PSL control for pure radar sensing. We adapt the algorithm in \cite{Li2008} by first solving $\min_{\mathbf{X} \in \mathcal{S}_{\text{X}}} \|\mathbf{Hx} - \mathbf{s}\|_{2}^2$ using the BSUM method and then following the approach in \cite{Li2008} to obtain the radar receive filter $\mathbf{F}$. We refer to the modified algorithm of \cite{Li2008} as the ``Exact Design'' scheme. Furthermore, we also implemented the ADMM algorithm and BSUM-based ALM algorithm with a fixed penalty parameter to show the superiority of the proposed adaptive penalty parameter scheme. We refer to them as ``ADMM'' and ``Fixed Penalty'' schemes, respectively. Without loss of generality, the penalty parameters for both the two algorithms are fixed to be $1$, and the total iterations of the ``ADMM'' algorithm are set the same as the proposed ALM algorithm. In the legends of the result figures, ``Direc'' refers to the receive beampattern focused at $-30^{\circ}$, $0^{\circ}$, and $30^{\circ}$ with a $3$ dB beamwidth of $10^{\circ}$, generated by the algorithm in \cite{li2007mimo}, and ``Omni'' denotes the omnidirectional desired beampattern, i.e., $\mathbf{R}_{\text{d}} = (P_{\text{X}}/N_{T})\mathbf{I}_{N_{T}}$.

Fig. \ref{fig:Performance_compare} presents the optimized results of different algorithms. In Fig. \ref{fig:Performance_compare}\subref{fig:rate_algs}, the proposed ALM algorithm achieves the highest sum rate under the directional beampattern case. For an omnidirectional beam, we need to set a larger $\alpha$ to make the final beampattern closer to omnidirectional because an omnidirectional beam requires that $\mathbf{X}$ and $\mathbf{F}$ be orthogonal in each row and column, which significantly shrinks the feasible region. However, this degrades the minimization on $P_{\text{MUI}}$ and leads to a lower achievable rate. In the Exact Design scheme, although the transmit sequence is optimized solely to minimize the MUI at the receiver, it still has a lower spectral efficiency compared to the result of the proposed ALM algorithm. This suggests that directly minimizing $\|\mathbf{H}\mathbf{x} - \mathbf{s}\|_2^2$ is prone to suboptimal local minima. In contrast, our proposed algorithm demonstrates a strong ability to escape such local minima and attain better solutions, even under more stringent constraints. Other methods, due to the inappropriate penalty parameter, provide lower achievable rates. Fig. \ref{fig:Performance_compare}\subref{fig:bpMSE_algs} shows that the Exact Design scheme nearly perfectly matches the desired beampatterns in both the directional and omnidirectional cases while the proposed inexact ALM algorithms have slight differences from the omnidirectional beampattern. The received beampatterns under the Fixed Penalty scheme and the ADMM scheme were far from the desired one because the inappropriate penalty parameter makes the algorithm focus too much on reducing the violation. Fig. \ref{fig:Performance_compare}\subref{fig:SL_algs} shows the proposed ALM algorithm has the best sidelobe control performance.

Fig. \ref{fig:images} shows the MIMO radar images formed by the results of different algorithms\footnote{We attempted to identify quantitative metrics for evaluating radar image quality, but to the best of our knowledge, no formal or widely accepted metric has been established. As in prior classical works \cite{Li2008, 4770164}, visual inspection remains the standard evaluation approach, which we also adopt in this study.}. The proposed inexact ALM algorithm achieved the best imaging performance. Although the Exact Design scheme perfectly meets the desired beampattern, it has a high sidelobe level, which results in severe interference caused by the clutter at the range bin of interest and, thus, a poorer MIMO radar image. Due to the inappropriate penalty parameter, the results under both the Fixed Penalty scheme and the ADMM scheme failed to produce radar images.
\section{Conclusion}\label{sec:conclusion}
This paper addressed the joint design of the receive filter and transmit waveform for MIMO-ISAC systems. We formulated an optimization problem to minimize a weighted sum of radar beampattern MSE and MUI at communication receivers, subject to practical constraints in convex and nonconvex forms. An inexact ALM algorithm was developed to solve the problem by iteratively minimizing tight upper bounds for each variable block. We proved the algorithm's convergence to a feasible stationary point, which is the best that one can expect for this optimization problem with many nonconvex constraints. The trade-offs between sum rate, beampattern MSE, sidelobe suppression size, and maximum sidelobe level were analyzed. Simulation results demonstrated that the proposed algorithm outperforms others across various metrics.
{\appendices
\section{Proof of Lemma 2}\label{proof:lemma2}
From \eqref{prob:block_aux_c}, it is clear that
\begin{multline}\label{eq:C_ineq}
  \mathcal{L}_{\rho^{(\ell)}}(\bm{C}^{(t)}, \mathbf{F}^{(t)}, \mathbf{X}^{(t)}; \bm{U}^{(\ell)}) - \mathcal{L}_{\rho^{(\ell)}}(\bm{C}^{(t+1)}, \mathbf{F}^{(t)}, \mathbf{X}^{(t)}; \bm{U}^{(\ell)}) \\
  \geq \frac{\beta}{2} \sum_{k \in \mathbf{\Omega}_{K}} \|\mathbf{C}_{k}^{(t)} - \mathbf{C}_{k}^{(t+1)}\|_{\text{F}}^2.
\end{multline}
It is simple to verify that updating $\mathbf{F}^{(t+1)}$ by solving problem \eqref{prob:PSL_F_MMed_lg0} is equivalent to
\begin{multline}\label{prob:prox_F}
  \mathbf{F}^{(t+1)} \in \argmin_{\mathbf{F} \in \mathbb{C}^{N_{\text{T}} \times L}} \{ \mathcal{R}\{\mathrm{tr}[\nabla_{\mathbf{F}} \mathcal{L}_{\text{F}}(\mathbf{F}^{(t)})^{\text{H}} (\mathbf{F} - \mathbf{F}^{(t)})]\} \\
  + \frac{\lambda_{q}^{(t)}}{2} \|\mathbf{F} - \mathbf{F}^{(t)}\|_{\text{F}}^2 + \mathbb{I}_{\mathcal{S}_{\text{F}}}(\mathbf{F})\},
\end{multline}
where $\lambda_{q}^{(t)}$ is defined in \eqref{eq:F_2_MM} that relies on $\mathbf{X}^{(t)}$. From Lemma 2 in \cite{bolte2014proximal}, we have
\begin{equation}
  \mathcal{L}_{\text{F}}(\mathbf{F}^{(t)}) - \mathcal{L}_{\text{F}}(\mathbf{F}^{(t+1)}) \geq \frac{\lambda_{q}^{(t)} - L_{f}^{(t)}}{2}\|\mathbf{F}^{(t+1)} - \mathbf{F}^{(t)}\|_{\text{F}}^2,\nonumber
\end{equation}
where $L_{f}^{(t)}$ denotes the Lipschitz constant of $\nabla_{\mathbf{F}}^{*} \mathcal{L}_{\text{F}}(\mathbf{F})$ at the $t$-th iteration of Algorithm \ref{alg:inexact_ALM_sol}, and $\nabla_{\mathbf{F}}^{*} \mathcal{L}_{\text{F}}(\mathbf{F})$ is the conjugate gradient of $\mathcal{L}_{\text{F}}(\mathbf{F})$. We always have $\lambda_{q}^{(t)} - L_{f}^{(t)} > 0$, since $L_{f}^{(t)}$ is the maximum eigenvalue of $\mathbf{Q}^{(t)}$ according to the results in \cite{zhou2018fenchel}, where $\mathbf{Q}^{(t)}$ is defined in \eqref{eq:F_2_MM} and its superscript $^{(t)}$ means it relies on $\mathbf{X}^{(t)}$. Then $\lambda_{q}^{(t)} > L_{f}^{(t)}$ for $\forall~t > 0$ according to \eqref{eq:F_2_MM}. Therefore, there exists a $\tau_{f} > 0$, such that the following holds for all $t > 0$:
\begin{multline}\label{eq:Ft_Ft1_ineq}
  \mathcal{L}_{\rho^{(\ell)}}(\mathbf{C}^{(t+1)}, \mathbf{F}^{(t)}, \mathbf{X}^{(t)}; \bm{U}^{(\ell)}) - \mathcal{L}_{\rho^{(\ell)}}(\mathbf{C}^{(t+1)}, \mathbf{F}^{(t+1)},   \\
  \mathbf{X}^{(t)}; \bm{U}^{(\ell)})\geq \tau_{f} \|\mathbf{F}^{(t)} - \mathbf{F}^{(t+1)}\|_{\text{F}}^2.
\end{multline}
By applying the above analysis procedure to \eqref{prob:X}, we obtain
\begin{multline}\label{eq:Xt_Xt1_ineq}
  \mathcal{L}_{\rho^{(\ell)}}(\bm{C}^{(t+1)}, \mathbf{F}^{(t+1)}, \mathbf{X}^{(t)}; \bm{U}^{(\ell)}) - \mathcal{L}_{\rho^{(\ell)}}(\bm{C}^{(t+1)}, \mathbf{F}^{(t+1)},                                                                                       \\
  \mathbf{X}^{(t+1)}; \bm{U}^{(\ell)})\geq \tau_{x} \|\mathbf{X}^{(\ell)} - \mathbf{X}^{(t+1)}\|_{\text{F}}^{2},
\end{multline}
where $\tau_{x} > 0$. Combining \eqref{eq:C_ineq}, \eqref{eq:Ft_Ft1_ineq}, and \eqref{eq:Xt_Xt1_ineq}, we have
\begin{multline}
  \mathcal{L}_{\rho^{(\ell)}}(\bm{z}^{(t)}; \bm{U}^{(\ell)}) - \mathcal{L}_{\rho^{(\ell)}}(\bm{z}^{(t+1)}; \bm{U}^{(\ell)})\geq \tau_{f} \|\mathbf{F}^{(t)} - \mathbf{F}^{(t+1)}\|_{\text{F}}^2 \\
  + \tau_{x} \|\mathbf{X}^{(t)} - \mathbf{X}^{(t+1)}\|_{\text{F}}^{2} + \frac{\beta}{2} \sum_{k \in \mathbf{\Omega}_{K}} \|\mathbf{C}_{k}^{(t)} - \mathbf{C}_{k}^{(t+1)}\|_{\text{F}}^2.
\end{multline}
Therefore, Lemma \ref{lm:2} holds.
\section{Proof of Lemma 3}\label{proof:lemma3}
\subsubsection{Upper Bound of $\mathbf{J}_{\rm F}^{(t+1)}$}
By solving problem \eqref{prob:PSL_F_MMed_lg0}, we have  $\mathbf{0} \in \nabla_{\mathbf{F}} \mathcal{G}_{\text{F}}(\mathbf{F}^{(t+1)}\mid\mathbf{F}^{(t)}) + \partial \mathbb{I}_{\mathcal{S}_{\text{F}}}(\mathbf{F}^{(t+1)})$. Hence, there exists a subgradient $\bm{A}^{(t+1)} \in \partial \mathbb{I}_{\mathcal{S}_{\text{F}}}(\mathbf{F}^{(t+1)})$ such that
\begin{equation}\label{eq:subgrad_GF}
  \nabla_{\mathbf{F}} \mathcal{G}_{\text{F}}(\mathbf{F}^{(t+1)}\mid\mathbf{F}^{(t)}) + \bm{A}^{(t+1)} = \mathbf{0},
\end{equation}
which further implies that there exists a subgradient $\mathbf{J}_{\text{F}}^{(t+1)} \in \partial_{\mathbf{F}} [\mathcal{L}_{\rho^{(\ell)}}(\bm{z}; \bm{U}^{(\ell)}) + \mathbb{I}_{\mathcal{S}_{\text{F}}}(\mathbf{F})] \mid _{\mathbf{F} = \mathbf{F}^{(t)}}$ such that
\begin{equation}\label{eq:subgrad_LF}
  \mathbf{J}_{\text{F}}^{(t+1)} = \nabla_{\mathbf{F}} \mathcal{L}_{\text{F}}(\mathbf{F}^{(t+1)}) + \bm{A}^{(t+1)}.
\end{equation}
By combining \eqref{eq:subgrad_GF} and \eqref{eq:subgrad_LF}, we have
\begin{multline}\label{eq:subgrad_LF_1}
  \|\mathbf{J}_{\text{F}}^{(t+1)}\|_{\text{F}}  = \|\nabla_{\mathbf{F}} \mathcal{L}_{\text{F}}(\mathbf{F}^{(t+1)}) - \nabla_{\mathbf{F}} \mathcal{G}_{\text{F}}(\mathbf{F}^{(t+1)}\mid\mathbf{F}^{(t)})\|_{\text{F}}                                \\
  \overset{(a)}{\leq} \|\nabla_{\mathbf{F}} \mathcal{L}_{\text{F}}(\mathbf{F}^{(t+1)}) - \nabla_{\mathbf{F}} \mathcal{L}_{\text{F}}(\mathbf{F}^{(t)})\|_{\text{F}} \\
  + \|\nabla_{\mathbf{F}} \mathcal{G}_{\text{F}}(\mathbf{F}^{(t+1)}\mid\mathbf{F}^{(t)}) - \nabla_{\mathbf{F}} \mathcal{G}_{\text{F}}(\mathbf{F}^{(t)}\mid\mathbf{F}^{(t)})\|_{\text{F}},
\end{multline}
where $(a)$ holds due to $\nabla_{\mathbf{F}} \mathcal{L}_{\text{F}}(\mathbf{F}^{(t)}) = \nabla_{\mathbf{F}} \mathcal{G}_{\text{F}}(\mathbf{F}^{(t)}\mid\mathbf{F}^{(t)})$. Since
\begin{multline}\label{ieq:L_F_f}
  \quad\|\nabla_{\mathbf{F}} \mathcal{L}_{\text{F}}(\mathbf{F}^{(t+1)}) - \nabla_{\mathbf{F}} \mathcal{L}_{\text{F}}(\mathbf{F}^{(t)})\|_{\text{F}}               \\
  = \|\mathbf{Q} (\mathbf{F}^{(t)} - \mathbf{F}^{(t+1)})\|_{\text{F}} \leq \|\mathbf{Q}\|_{\text{F}} \|\mathbf{F}^{(t)} - \mathbf{F}^{(t+1)}\|_{\text{F}},
\end{multline}
and
\begin{multline}\label{ieq:G_F_f}
  \|\nabla_{\mathbf{F}} \mathcal{G}_{\text{F}}(\mathbf{F}^{(t+1)}\mid\mathbf{F}^{(t)}) - \nabla_{\mathbf{F}} \mathcal{G}_{\text{F}}(\mathbf{F}^{(t)}\mid\mathbf{F}^{(t)})\|_{\text{F}} \\
  = \lambda_{q}\|\mathbf{F}^{(t)} - \mathbf{F}^{(t+1)}\|_{\text{F}} \leq \|\mathbf{Q}\|_{\text{F}} \|\mathbf{F}^{(t)} - \mathbf{F}^{(t+1)}\|_{\text{F}},
\end{multline}
it follows that
\begin{equation}\label{eq:sub_grad_bound_F}
  \|\mathbf{J}_{\text{F}}^{(t+1)}\|_{\text{F}} \leq \bar{L}_{\text{F}} \|\mathbf{F}^{(t)} - \mathbf{F}^{(t+1)}\|_{\text{F}},
\end{equation}
where the inequality holds due to
\begin{multline}
  \|\mathbf{Q}\|_{\text{F}} \leq \alpha \|\mathbf{X}\|_{\text{F}} + \sum_{k \in \mathbf{\Omega}_{K}} \frac{\rho^{(\ell)}}{2} \|\mathbf{X} \mathbf{J}_{k}\|_{\text{F}} \\
  \leq \sqrt{P_{\text{X}}} \left(\alpha + \frac{1}{2}\rho^{(\ell)} \sqrt{T} |\mathbf{\Omega}_{K}|\right) = \frac{\bar{L}_{\text{F}}}{2}. \nonumber
\end{multline}
\subsubsection{Upper Bound of $\mathbf{J}_{\rm X}^{(t+1)}$}
By following the similar analysis procedure in the previous part, we have
\begin{equation}\label{eq:sub_grad_bound_X}
  \|\mathbf{J}_{\text{X}}^{(t+1)}\|_{\text{F}} \leq \bar{L}_{\text{X}} \|\mathbf{X}^{(t)} - \mathbf{X}^{(t+1)}\|_{\text{F}},
\end{equation}
where $\bar{L}_{\text{X}} = 2 [\sqrt{P_{\text{F}}} (\alpha + \frac{1}{2} \rho^{(\ell)} \sqrt{T} |\mathbf{\Omega}_{K}|) + (1-\alpha)\|\mathbf{H}^{\text{H}}\mathbf{H}\|_{\text{F}}]$.

\subsubsection{Upper Bound of $\mathbf{J}_{C}^{(t+1)}$}
We now consider calculating the upper bound of the subgradient vector
\begin{equation}
  \mathbf{J}_{C}^{(t+1)} = \left[(\mathbf{J}_{C_{1}}^{(t+1)})^{\text{T}}, (\mathbf{J}_{C_{2}}^{(t+1)})^{\text{T}}, \dots, (\mathbf{J}_{C_{|\mathbf{\Omega}_{K}|}}^{(t+1)})^{\text{T}}\right]^{\text{T}}, \nonumber
\end{equation}
where $\mathbf{J}_{C_{k}}^{(t+1)}$ denotes the subgradient of $\mathcal{L}_{\rho^{(\ell)}}(\bm{z}; \bm{U}^{(\ell)}) + \mathbb{I}_{\mathcal{S}_{\text{C}}}(\mathbf{C}_{k})$ w.r.t. $\mathbf{C}_{k}$ at $\mathbf{C}_{k}^{(t+1)}$. Denoting the objective function in \eqref{prob:block_aux_c} as $\mathcal{H}_{\rho^{(\ell)}, k}(\mathbf{C}, \mathbf{F}^{(t)}, \mathbf{X}^{(t)})$ for each $k \in \mathbf{\Omega}_{K}$, we have
\begin{equation}\label{eq:subgrad_Ck}
  \mathbf{0} = \nabla_{\mathbf{C}_{k}} \mathcal{H}_{\rho^{(\ell)}, k}(\mathbf{C}_{k}^{(t+1)}, \mathbf{F}^{(t)}, \mathbf{X}^{(t)}) + \bm{D}_{k}^{(t+1)},
\end{equation}
where $\bm{D}_{k}^{(t+1)} \in \partial \mathbb{I}_{\mathcal{S}_{\text{C}}}(\mathbf{C}_{k}^{(t+1)})$. Then there exists a subgradient such that
$$
  \mathbf{J}_{C_{k}}^{(t+1)} = \nabla_{\mathbf{C}_{k}} \mathcal{L}_{\rho^{(\ell)}}(\bm{z};\bm{U}^{(\ell)}) \mid _{\bm{z} = \bm{z}^{(t+1)}} + \bm{D}_{k}^{(t+1)}.
$$

Following the similar derivation procedure regarding $\mathbf{J}_{\text{F}}^{(t+1)}$, we have the upper bound of $\|\mathbf{J}_{C_{k}}\|_{\text{F}}$. By summing $\mathbf{J}_{C_{k}}^{(t+1)}$ over $k \in \mathbf{\Omega}_{K}$, we finally get
\small\begin{multline}\label{eq:sub_grad_bound_C}
  \|\mathbf{J}_{C}^{(t+1)}\|_{\text{F}} \leq \frac{1}{2} \rho^{(\ell)} \sqrt{T} |\mathbf{\Omega}_{K}| \left(\sqrt{P_{\text{F}}} \| (\mathbf{X}^{(t)} - \mathbf{X}^{(t+1)}) \|_{\text{F}}  \right.                          \\
  \left. + \sqrt{P_{\text{X}}}\|\mathbf{F}^{(t)} - \mathbf{F}^{(t+1)}\|_{\text{F}}\right)+ \beta \sum_{k \in \mathbf{\Omega}_{K}} \| (\mathbf{C}_{k}^{(t+1)} - \mathbf{C}_{k}^{(t)}) \|_{\text{F}}.
\end{multline}\normalsize
Combining \eqref{eq:sub_grad_bound_F}, \eqref{eq:sub_grad_bound_X}, and \eqref{eq:sub_grad_bound_C} together, we arrive at Lemma \ref{lm:3}.
\section{Proof of Theorem 1}\label{proof:thm1}
From the results of Lemma \ref{lm:2} and Lemma \ref{lm:3}, with penalty parameter $\rho^{(\ell)}$ at the $\ell$-th iteration of Algorithm \ref{alg:RALM}, there exists a positive real number $M>0$ such that
\begin{multline}\label{eq:ALF_subgrad}
  M [\mathcal{L}_{\rho^{(\ell)}}(\bm{z}^{(t)}; \bm{U}^{(\ell)}) - \mathcal{L}_{\rho^{(\ell)}}(\bm{z}^{(t+1)}; \bm{U}^{(\ell)})]  \geq M [\|\mathbf{X}^{(t)}                                       \\
  -    \mathbf{X}^{(t+1)}\|_{\text{F}}^2 + \|\mathbf{F}^{(t)} - \mathbf{F}^{(t+1)}\|_{\text{F}}^2 + \frac{\beta}{2} \sum_{k \in \mathbf{\Omega}_{K}} \|\mathbf{C}_{k}^{(t)}- \mathbf{C}_{k}^{(t+1)}\|_{\text{F}}^2]        \\
  \geq \frac{1}{{\rho^{(\ell)}}^2} (\|\mathbf{J}_{\text{F}}^{(t+1)}\|_{\text{F}}^2  + \|\mathbf{J}_{\text{X}}^{(t+1)}\|_{\text{F}}^2 + \|\mathbf{J}_{C}^{(t+1)}\|_{\text{F}}^2) = \frac{\|\bm{J}^{(t+1)}\|_{\text{F}}^{2}}{{\rho^{(\ell)}}^2},
\end{multline}
where $\bm{J}^{(t+1)}$ is defined in Lemma \ref{lm:3}. The sufficient descent property in Lemma \ref{lm:3} guarantees that the left-hand side of \eqref{eq:ALF_subgrad} is always nonnegative. By summing \eqref{eq:ALF_subgrad} from $1$ to $T$, we have
\begin{multline}\label{eq:subgrad_and_T}
  M {\rho^{(\ell)}}^2 [\mathcal{L}_{\rho^{(\ell)}}(\bm{z}^{(1)}; \bm{U}^{(\ell)}) - \mathcal{L}_{\rho^{(\ell)}}(\bm{z}^{(T)}; \bm{U}^{(\ell)})] \\
  \geq \sum_{t =1}^{T} \|\bm{J}^{(t)}\|_{\text{F}}^{2} \geq T \min_{t \in [1, T]} \|\bm{J}^{(t)}\|_{\text{F}}^{2}.
\end{multline}
The value of ALF is lower bounded, i.e., $\mathcal{L}_{\rho}(\bm{z}; \bm{U}) > -\infty$, since the summation of Frobenius norms in the ALF is no less than zero and the term $\|\mathbf{U}_{k}\|_{\text{F}}^2/(2\rho^{(\ell)})$ is finite for all $k \in \mathbf{\Omega}_{K}$ due to the boundness of multipliers. The $\varepsilon^{(\ell)}$-stationary point means $\min_{t \in [1, T]} \|\bm{J}^{(t)}\|_{\text{F}} \leq \varepsilon^{(\ell)}$. Then, \eqref{eq:subgrad_and_T} yields that
\begin{equation}
  T \leq \frac{M u {\rho^{(\ell)}}^2}{{\varepsilon^{(\ell)}}^2}, \nonumber
\end{equation}
where $u = \mathcal{L}_{\rho^{(\ell)}}(\bm{z}^{(1)}; \bm{U}^{(\ell)}) - \mathcal{L}_{\rho^{(\ell)}}(\bm{z}^{(T)}; \bm{U}^{(\ell)})$ is a positive number. Therefore, the sequences generated by Algorithm \ref{alg:inexact_ALM_sol} will achieve an $\varepsilon^{(\ell)}$-stationary point after $\mathcal{O}\left(\frac{{\rho^{(\ell)}}^2}{{\varepsilon^{(\ell)}}^2}\right)$ iterations.
\section{Proof of Theorem 2}\label{proof:thm2}
The convergence proof consists of two main steps, where the first step shows the feasibility of the limit point of the sequence generated by Algorithm 1 and the second step shows that any limit point is a feasible stationary point of problem \eqref{prob:prob_origin}. We present these two steps in two subsections separately.
\subsection{Feasibility of the Limit Point}
According to the updating rule of the penalty parameter in \eqref{eq:rho_update}, either the penalty parameter keeps fixed after a finite number of iterations or $\lim_{\ell \rightarrow \infty} \rho^{(\ell+1)} = +\infty$. The violation of the constraints will be decreased by a factor of $\delta < 1$ after each iteration and thus tends to zero as $\ell \rightarrow \infty$ if it is the former case.

Now we consider the condition where $\rho^{(\ell)} \rightarrow \infty$. Denote $(\bm{C}^{(\infty)}, \mathbf{F}^{(\infty)}, \mathbf{X}^{(\infty)})$ as any limit point of the sequence generated by Algorithm \ref{alg:RALM}. Based on the choice of the initial point in \eqref{eq:init_point_at_lth_ALg1} and the previous analysis, we can see that condition \eqref{ineq:upper_bound_requirement} holds for any $\ell \geq 0$. That is
\begin{multline}
  f^{(\ell+1)} + \sum_{k \in \mathbf{\Omega}_{K}} \Big[\frac{\rho^{(\ell)}}{2} \|\mathbf{X}^{(\ell+1)} \mathbf{J}_{k} {\mathbf{F}^{(\ell+1)}}^{\text{H}} - \mathbf{C}_{k}^{(\ell+1)}\|_{\text{F}}^2                 \\
  + \mathcal{R}\Big\{tr\Big[{\mathbf{U}_{k}^{(\ell)}}^{\text{H}} \Big(\mathbf{X}^{(\ell+1)} \mathbf{J}_{k} {\mathbf{F}^{(\ell+1)}}^{\text{H}} - \mathbf{C}_{k}^{(\ell+1)}\Big)\Big]\Big\}\Big] \leq \zeta,\nonumber
\end{multline}
where $f^{(\ell+1)} = f(\mathbf{F}^{(\ell+1)}, \mathbf{X}^{(\ell+1)})$. Dividing both sides of the above inequality by $\rho^{(\ell)}$ yields
\begin{multline}
  \sum_{k \in \mathbf{\Omega}_{K}} \frac{1}{2} \|\mathbf{X}^{(\ell+1)} \mathbf{J}_{k} {\mathbf{F}^{(\ell+1)}}^{\text{H}} - \mathbf{C}_{k}^{(\ell+1)}\|_{\text{F}}^2 \leq \frac{1}{\rho^{(\ell)}} \Big[\zeta - f^{(\ell+1)}                 \\
    - \sum_{k \in \mathbf{\Omega}_{K}} \mathcal{R}\Big\{tr\Big[{\mathbf{U}_{k}^{(\ell)}}^{\text{H}} \Big(\mathbf{X}^{(\ell+1)} \mathbf{J}_{k} {\mathbf{F}^{(\ell+1)}}^{\text{H}} - \mathbf{C}_{k}^{(\ell+1)}\Big)\Big]\Big\} \Big]. \nonumber
\end{multline}
Since $\zeta$ is a finite constant given in \eqref{eq:zeta_the_upper_bound}, $f(\mathbf{F}, \mathbf{X})$ is lower bounded by zero, and $\mathbf{U}_{k}$ is also bounded according to \eqref{eq:uk_update}, it follows that the right-hand side of the above inequality tends to zero when $\rho^{(\ell)} \rightarrow \infty$. Therefore,
\begin{equation}
  \lim_{\ell \rightarrow \infty} \sum_{k \in \mathbf{\Omega}_{K}} \|\mathbf{X}^{(\ell+1)} \mathbf{J}_{k} {\mathbf{F}^{(\ell+1)}}^{\text{H}} - \mathbf{C}_{k}^{(\ell+1)}\|_{\text{F}}^2 \leq 0, \nonumber
\end{equation}
which implies $\|\mathbf{X}^{(\infty)} \mathbf{J}_{k} {\mathbf{F}^{(\infty)}}^{\text{H}} - \mathbf{C}_{k}^{(\infty)}\|_{\text{F}} = 0$ for all $k \in \mathbf{\Omega}_{K}$. Therefore, any limit point of the sequence generated by Algorithm \ref{alg:RALM} is feasible.
\subsection{The Limit Point is a Stationary Feasible Solution of Problem \eqref{prob:prob_origin}}}
Based on Definition \ref{def_1} and Theorem \ref{thm:1}, we know that as $\varepsilon^{(\ell)} \to 0$ and $\ell \to \infty$, any limit point generated by Algorithm \ref{alg:RALM} is a stationary point. In the previous subsection, we have shown that every such limit point is also a feasible solution to problem \eqref{prob:FX_with_C}. Hence, the pair $(\mathbf{F}^{(\infty)}, \mathbf{X}^{(\infty)})$ constitutes a feasible stationary point of problem \eqref{prob:FX_with_C}. Furthermore, according to \cite[Proposition 1]{Bolte2018}, any feasible stationary point of the augmented Lagrangian problem is also a feasible stationary point of the original problem. Therefore, we conclude that every limit point generated by Algorithm \ref{alg:RALM} is a feasible stationary solution of the original problem \eqref{prob:prob_origin}.
%%%%%%%%%%%%%%%%%%%%%%%%%%%%%%%%%%%%%%%%%%%%%%%%%%%%%%%%%%%%%%%%%%%%%%
\bibliographystyle{IEEEtran}
\bibliography{refs}
\vfill
\end{document}